\documentclass[conference]{IEEEtran}

\usepackage{tikz}
\usepackage{mathptmx}
\usepackage[normalem]{ulem}
\usepackage{multicol}
\usepackage{pifont}
\usepackage{multirow}
\usepackage{color}
\usepackage{listings}
\usepackage{float}
\usepackage{ragged2e}
\usepackage{xspace}
\usepackage{soul}
\usepackage{lipsum}
\usepackage{makecell}
\usepackage{mathtools}
\usepackage{subcaption}
\usepackage[final]{microtype}
\usepackage{flushend}
\usepackage{listings}
\usepackage{multirow}
\usepackage{wrapfig}
\usepackage{etoolbox}
\usepackage{amsmath,amssymb,amsfonts}
\usepackage{algorithmic}
\usepackage{graphicx}
\usepackage{textcomp}
\usepackage{xcolor}
\usepackage[ruled,vlined,linesnumbered,algo2e]{algorithm2e}

\usepackage{subcaption}

\usepackage{enumitem}

\usepackage{cite}
\usepackage{amsmath,amssymb,amsfonts}
\usepackage{algorithmic}
\usepackage{graphicx}
\usepackage{textcomp}
\usepackage{xcolor}
\usepackage[hyphens]{url}
\usepackage{fancyhdr}
\usepackage{hyperref}

\newcommand{\proj}{\benchmark{MANT}\xspace}
\newcommand{\Fig}[1]{Fig.~\ref{#1}}

\newcommand{\Tbl}[1]{Tbl.~\ref{#1}}
\newcommand{\Sec}[1]{Sec.~\ref{#1}}

\newcommand{\benchmark}[1]{{\texttt{#1}}}
\renewcommand{\paragraph}[1]{\vspace*{0.15cm}\noindent\textbf{#1}\hspace*{.1cm}}

\newcommand{\ra}[1]{\renewcommand{\arraystretch}{#1}}

\newcommand\blankfootnote[1]{%
            \let\thefootnote\relax\footnotetext{#1}%
            \let\thefootnote\svthefootnote%
          }

\definecolor{myyellow}{RGB}{254, 230, 153}
\definecolor{myblue}{RGB}{193, 228, 248}
\definecolor{mygreen}{RGB}{24, 154, 76}

\def\BibTeX{{\rm B\kern-.05em{\sc i\kern-.025em b}\kern-.08em
    T\kern-.1667em\lower.7ex\hbox{E}\kern-.125emX}}

\newcommand{\hpcayear}{2025}

\title{M-ANT: Efficient Low-bit Group Quantization for LLMs via Mathematically Adaptive Numerical Type 
\thanks{\IEEEauthorrefmark{1} Jingwen Leng is corresponding authors of this paper.}
}

\author{
  \IEEEauthorblockN{
    Weiming Hu$^{1, 2}$,
    Haoyan Zhang$^{1, 2}$,
    Cong Guo$^3$, 
    Yu Feng$^{1}$,
    Renyang Guan$^{1, 2}$,
    Zhendong Hua$^{1}$,
    Zihan Liu$^{1, 2}$, \\
    Yue Guan$^{1, 2}$, 
    Minyi Guo$^{1, 2}$,
    Jingwen Leng$^{1, 2,}$\IEEEauthorrefmark{1}
  }
  \IEEEauthorblockA{
    $^1$Shanghai Jiao Tong University, $^2$Shanghai Qi Zhi Institute, $^3$Duke University \\
    \{weiminghu, h.y.zhang-zdy, y-feng, guanrenyang, jackeyhua, altair.liu, bonboru\}@sjtu.edu.cn, \\ \{guo-my, leng-jw\}@cs.sjtu.edu.cn, cong.guo@duke.edu
  }

}

\def\hpcacameraready{} 

\fancypagestyle{camerareadyfirstpage}{%
  \fancyhead{}
  
  \fancyhead[C]{
    \ifdefined\aeopen
    \parbox[][12mm][t]{13.5cm}{\hpcayear{} IEEE International Symposium on High-Performance Computer Architecture (HPCA)}    
    \else
      \ifdefined\aereviewed
      \parbox[][12mm][t]{13.5cm}{\hpcayear{} IEEE International Symposium on High-Performance Computer Architecture (HPCA)}
      \else
      \ifdefined\aereproduced
      \parbox[][12mm][t]{13.5cm}{\hpcayear{} IEEE International Symposium on High-Performance Computer Architecture (HPCA)}
      \else
      \parbox[][0mm][t]{13.5cm}{\hpcayear{} IEEE International Symposium on High-Performance Computer Architecture (HPCA)}
    \fi 
    \fi 
    \fi 
    \ifdefined\aeopen 
      \includegraphics[width=12mm,height=12mm]{ae-badges/open-research-objects.pdf}
    \fi 
    \ifdefined\aereviewed
      \includegraphics[width=12mm,height=12mm]{ae-badges/research-objects-reviewed.pdf}
    \fi 
    \ifdefined\aereproduced
      \includegraphics[width=12mm,height=12mm]{ae-badges/results-reproduced.pdf}
    \fi
  }
  \fancyfoot[C]{}
}
\begin{document}

\maketitle
\pagestyle{plain}

\ifdefined\hpcacameraready 
  \thispagestyle{camerareadyfirstpage}
  \pagestyle{empty}
\else
  \thispagestyle{plain}
  \pagestyle{plain}
\fi

\newcommand{\hpcaheight}{0mm}
\ifdefined\eaopen
\renewcommand{\hpcaheight}{12mm}
\fi

\begin{abstract} 
Large language models (LLMs) are one of the most important killer computer applications. 
The recent algorithmic advancement proposes a fine-grained group-wise quantization for LLMs, which treats a small set (e.g., 64) of values in a tensor as a compression unit.
It effectively preserves the model accuracy without retraining, and has become the standard approach to efficiently deploy LLMs.
On the other hand, there are works that propose various adaptive data types to better adapt to different distributions and further reduce the required bit length for LLMs.
In this work, our detailed analysis unveils a key finding that while different tensors exhibit similar distributions, small groups can have markedly different distributions.
As such, the group-level diversity requires a new level of adaptivity for which existing adaptive data types fail to provide.




In this paper, we propose \proj, a \underline{m}athematically \underline{a}daptive \underline{n}umeric \underline{t}ype, featuring a more flexible encoding paradigm with a wider range of data distribution and more efficient decoding-computation fusion mechanism to address these challenges. 
Based on \proj, we develop a supporting framework to assign the appropriate data type for each group adaptively. 
Meanwhile, the dynamically generated Key-Value (KV) caches in LLMs introduce further complexity for real-time quantization.
To tackle this, we propose an efficient real-time quantization mechanism. 
Besides, we implement a specific processing element (PE) to efficiently support \proj and incorporate a real-time quantization unit. 
By integrating these components into a systolic array, \proj unifies the group-wise weight and KV cache quantization and addresses the associated challenges. 
Our evaluation shows achieving, on average, 2.99$\times$ (up to 4.46$\times$) speedup and 2.81$\times$ (up to 4.10$\times$) energy reduction to the state-of-the-art LLM accelerator.

\end{abstract}

\section{Introduction}\label{sec:introduction}
\blankfootnote{* Jingwen Leng is the corresponding author of this paper.}
In recent years, Large Language Models (LLMs) have demonstrated significant improvements in quality and accuracy across various natural language processing (NLP) tasks~\cite{chowdhery2022palm, scao2022bloom, touvron2023llama, zhang2022opt}.
However, the rapid increase in LLM parameters poses substantial challenges to memory and computational resources on existing hardware platforms like GPUs~\cite{v100,a100,h100} and TPUs~\cite{jouppi2023tpu}.
The latest Llama3 model~\cite{dubey2024llama3herdmodels}, with over 405 billion parameters, requires about 800 GB of memory, exceeding the capabilities of high-end hardware platforms such as the H100 GPU~\cite{h100}, which has 80~GB memory.
Additionally, single-batch inference involves sequentially predicting each token based on the previous ones due to the auto-regressive nature of LLMs.
As a result, LLM inference mainly depends on narrow matrix multiplications, making it memory-bound and leading to severe under-utilization of GPUs' computation resources.

Quantization has emerged as a promising method to mitigate LLMs' challenges in memory consumption and inference latency, which can be applied at different granularities such as the level of a channel~\cite{xiao2023smoothquant,lee2024tenderacceleratinglargelanguage,kim2023squeezellm} and a tensor~\cite{guo2022ant,zadeh2022mokey,zadeh2020gobo}.
Recent studies advocate for group-wise quantization, which uses a group, such as 64 contiguous elements within a channel, as the unit of quantization granularity~\cite{zhao2023atom,lin2023awq,frantar2023gptq,shao2024omniquant,dai2021vsquant,liu2024kivi}.
The group quantization reduces the quantization error with negligible software overhead, and hence has become the standard method for accelerating LLMs (\Sec{sec:acc_analysis}).

The above quantization methods mostly use \texttt{INT} or \texttt{FP} data types, and there are works that propose various adaptive data types to better adapt to different distributions and hence improve the quantized model performance.
The adaptive data type method falls into two main categories: data-type-based and clustering-based.
The former category selects data types from discrete sets based on tensor data distribution.
The representative work ANT~\cite{guo2022ant} packages several different data types for selection, including \texttt{INT} for the uniform distribution, \texttt{PoT} (Power of Two) for the Laplace distribution, and \texttt{flint} for the Gaussian distribution.
The latter category utilizes clustering algorithms to generate centroids that align with the data distribution and provide considerable adaptivity. 
Mokey~\cite{zadeh2022mokey} and GOBO~\cite{zadeh2020gobo} exemplify this approach (\Sec{sec:acc_analysis}).
 


However, our analysis shows that those existing adaptive methods fail to achieve optimal performance in terms of accuracy and resource efficiency.
While different tensors exhibit similar distributions, small groups can have markedly different distributions. 
This finding underscores the necessity for full adaptivity in group quantization to fully realize its potential.
Using the per-group clustering as the ideal baseline, we show that existing data-type-based adaptive method ANT~\cite{guo2022ant} still suffers from a significant accuracy loss (\Sec{sec:acc_analysis}).

We also show that it is not feasible to transform existing adaptive methods for the group quantization paradigm.
For the adaptive data type methods, a naive approach to enhance adaptivity is to expand their set of data types. However, incorporating new data types requires additional decoders, which increase hardware costs. 
Additionally, compatibility issues between new and existing data types may reduce computational efficiency. For instance, the \texttt{NF4} data type~\cite{dettmers2023qlora} requires an FP16 MAC unit, which is incompatible with existing \texttt{ANT} data types.
Clustering-based adaptive methods like GOBO~\cite{zadeh2020gobo} and Mokey~\cite{zadeh2022mokey} can sufficiently adapt to various distributions at the group level. However, they require codebooks for quantization and dequantization, leading to high adaptivity at the expense of encoding and computational efficiency. 
For instance, GOBO~\cite{zadeh2020gobo} employs the K-means algorithm to quantize weights and requires dequantization to \texttt{FP16} using a codebook lookup table before computation, resulting in high adaptivity but low computational efficiency (\Sec{subsec:efficiency}). 

Finally, it is also challenging to satisfy the real-time quantization of LLMs' KV caches, which can occupy over $70\%$ of memory and computation resources when the sequence length is sufficiently long~\cite{hooper2024kvquant,zhao2023atom,sheng2023flexgen}. 
As a result, the efforts on weight and activation quantization become marginal due to this dominance of KV cache.
Although, in the LLM inference, KV cache behaves more like dynamically generated ``weights" for the $Q$ values in the attention mechanism~\cite{vaswani2017attention},
quantizing the KV cache at a fine-grained granularity is challenging due to substantial runtime overhead. 
Most previous quantization efforts~\cite{lin2023awq,guo2022ant,guo2023olive,frantar2023gptq,dettmers2023spqr,dettmers2023qlora}, encompassing weight-activation and weight-only quantization have sidestepped the issue of KV cache quantization (\Sec{sec:moti_kvcache}).

In this study, we introduce \proj{}, \underline{m}athematically \underline{a}daptive \underline{n}umeric \underline{t}ype, a novel adaptive method that aims at overcoming the above challenges, which can be summarized into three points: decoding and computation efficiency, group-wise diversity, and KV cache dynamics.
To tackle the first two challenges related to efficiency and diversity, we propose \proj{} with a simple yet mathematical formulation to support various data types. 
The formulation of \proj{} is designed to encompass a wide spectrum of data types, theoretically extending to ``infinite'' variations through a smooth transition in the data type distribution. 
Examples of supported data types include \texttt{NormalFloat}~\cite{dettmers2023qlora} and \texttt{flint}~\cite{guo2022ant}.
Moreover, adopting a group-wise design approach, we enable \proj{} to operate at a fine-grained granularity.
This inherently addresses the diversity and adaptivity requirements within LLMs. 
Employing group-wise formulated data types introduces computational overhead due to the diversity of data types at this fine-grained granularity.
Thus, we introduce a novel computing paradigm named \proj, to mitigate this overhead significantly (\Sec{sec:encode}).

To address the KV cache's dynamic challenge, we introduce a novel real-time quantization method, which lets us integrate weight and KV cache quantization into a unified framework.
In our approach, we regard the KV cache as ``dynamic weights'', as opposed to the ``static weights'' of the original weights.
Despite having identical tensor shapes, the K cache and V cache employ distinct computation methods, significantly complicating our design.
We propose two real-time quantization techniques in two dimensions: spatial and temporal. 
Spatial quantization is tailored for the K cache, where all elements within a K cache group are generated simultaneously. 
To accommodate this, we allocate sufficient quantization processing units for the K cache.
Conversely, the V cache presents a more intricate challenge for real-time quantization. In the V cache, elements within a group are generated across the temporal dimension, meaning they emerge in different continuous iterations. 
Consequently, we propose a highly efficient temporal real-time quantization method tailored to the V cache (\Sec{sec:dse}).

Finally, \proj proposes efficient group-wise quantization for both weight and KV cache.
We augment the processing element (PE) in the classic systolic array architecture to support our encoding paradigm and implement a real-time quantization engine to accelerate KV cache quantization (\Sec{sec:architecture}).
We show that the components of \proj introduce negligible area overhead with significant speedup and energy reduction compared to the Transformer accelerators~\cite{guo2022ant,zadeh2022mokey,guo2023olive,lee2024tenderacceleratinglargelanguage}.

In summary, this paper makes the following contributions.

\begin{itemize}[leftmargin=*]
	\item We analyze the challenges of group-wise quantization in both weight and KV cache, and design a new encoding paradigm to improve information utilization in group-wise quantization and inherently enable efficient decoding in LLM inference.
    \item We introduce a specific processing element (PE) to provide efficient encoding and computation for our encoding paradigm, and by integrating the real-time quantization components for KV cache, we implement a unified quantization system for group-wise quantization in weight and KV cache.
    \item We integrate these components into an existing systolic array and achieve up to 4.46$\times$ speedup and 4.10$\times$ energy reduction compared to existing works, with negligible area overhead.
\end{itemize}
\section{Background and Related Work}
\label{sec:background}


In this section, we first present the background of LLMs and their inference processes.
Next, we examine quantization techniques, emphasizing group-wise quantization and commonly used data types in quantization.

\subsection{LLM Inference Process}

LLMs often adopt the Transformer architecture~\cite{vaswani2017attention} in LLMs.
Each Transformer layer incorporates an attention module and a feed-forward network (FFN).
Mathematically, multi-head attention can be described as follows:
\begin{align*}
\text{MultiHead}(Q, K, V) &= \text{Concat}(\text{head}_1, \dots, \text{head}_h)W^O, \\
\text{head}_i &= \text{Attention}(HW_i^Q, HW_i^K, HW_i^V), \\
\text{Attention}(Q, K, V) &= \text{softmax}(QK^T/\sqrt{d_k})V.
\end{align*}
Here, $W_i^Q$, $W_i^K$, $W_i^V$, and $W^O$ are parameter matrices for the $i$-th head and the output projection, respectively. And $H$ is the hidden state. The $softmax$ function is applied over the keys to normalize their weights, ensuring that the output is a weighted sum of the values based on the input's relevance.

The generative inference process of LLM typically involves two distinct stages: \textbf{prefill} stage and \textbf{decode} stage. 
The prefill stage supplies a prompt sequence to the model, establishing the context for subsequent text generation.
Simultaneously, the attention operation produces $K$ and $V$ tensors, which are cached using the \textbf{KV cache} technique to prevent redundant recomputation during the subsequent decode stage.
The decode stage is where actual text generation occurs.
In this stage, the model uses each generated token along with the KV cache to iteratively generate the next token and update the KV cache.
The KV cache continues to expand until the process meets a specific termination condition.
Since the \textbf{decode} stage requires only a single token as input, its operations primarily consist of matrix-vector multiplication (GEMV), resulting in a memory-bound scenario dominated by memory bandwidth.

The memory footprint of generative LLM inference comprises two major components: the weight and the KV cache, either of which can be the memory bottleneck under various scenarios~\cite{hooper2024kvquant, liu2024kivi,kang2024gear,zhao2023atom,sheng2023flexgen,guo2024survey}.
Compression techniques such as quantization~\cite{han2015deep,guo2022ant,zadeh2022mokey,zadeh2020gobo,zhao2023atom,lin2023awq,guo2023olive, frantar2023gptq,li2023efficientadaptiveactivationrounding,dettmers2022llm} and sparsity~\cite{guo2020accelerating,frantar2023sparsegptmassivelanguagemodels,wang2021dual,guan2020far,guan2022block,guan2022transkimmer,guan2024fractal,NIPS1989_6c9882bb,zhang_h_2o_2023,zhang_q-hitter_2024,zhang2024dstc, guo2024accelerating,qiu2019adversarial} emerge to solve these issue.

\subsection{Quantization Technique}
\label{sec:bg_quantization}

The quantization technique~\cite{yao2020zeroquant,dettmers2022llm,shao2024omniquant,frantar2023gptq,guo2022squant,shen2020q} is an effective method for compressing neural network parameters with minimal loss.
The equation below defines the quantized data $W'$ and the dequantized data $\hat{W}$, where $s$ represents the scaling factor determined by the ranges of the two data types:
\begin{equation}
    \begin{array}{c}
        W' = \lfloor \frac{W}{s} \rceil, \hat{W} = s \times W' \\
    \end{array}
\label{quant_and_deq}
\end{equation}
Traditional quantization methods assign mappings at the tensor or channel level, known as tensor- or channel-wise quantization.
In these cases, outliers can significantly affect quantization performance by greatly increasing the rounding errors of standard values across the entire tensor or channel~\cite{dettmers2022llm,guo2023olive}.

\paragraph{Group-wise Quantization.}
To address the outlier challenge, various studies~\cite{zhao2023atom,lin2023awq,frantar2023gptq,shao2024omniquant,dai2021vsquant} advocate for group-wise quantization, which uses a group, such as 64 contiguous elements within a channel, as the unit of quantization granularity.
Consider a tensor with dimensions (2048, 4096), comprising 4096 channels.
With a group size of 128, each channel contains $(4096/128) = 32$ groups, resulting in a total of $2048 \times 32 = 65536$ groups across the entire tensor.
Despite the modest overhead from the mapping parameter for each group, fine-grained group-wise quantization limits the impact of outliers to smaller regions, thereby enhancing performance.

\paragraph{Data Type for Quantization.}
Several studies~\cite{guo2022ant,dettmers2023qlora,zadeh2022mokey,ramachandran2024algorithmhardware} adapt to the different distribution observed at tensor or channel levels by using customized data types. ANT~\cite{guo2022ant} introduces \texttt{flint} and employs an adaptive method to select the most suitable numerical data type for each tensor from a predefined set of data types. Building on ANT, OliVe~\cite{guo2023olive} develops \texttt{abfloat} to more accurately represent the distribution of outliers. Both ANT and OliVe enhance their systems with custom decoders and MAC (multiply-accumulate) units to facilitate their specific arithmetic computation workflows. Moreover, QLoRA~\cite{dettmers2023qlora} introduces \texttt{NormalFloat} (\texttt{NF}), which is derived from Gaussian distribution's quantile data points, for a more precise fit. Unlike other methods, \texttt{NF} requires high-precision dequantization before any computational operations, as it does not directly support multiplication and accumulation. In contrast, Mokey~\cite{zadeh2022mokey} uses clustering to derive the best date type for the quantization unit.
However, it requires additional codebooks to store cluster centroids.
Thus, it proposes \texttt{golden dictionary} (\texttt{GD}) to mitigate those overheads.
Microscaling \texttt{float} (\texttt{MXFP})~\cite{2023mxfp,NEURIPS2020_747e32ab} combines block-wise shared scale with \texttt{float} and \texttt{INT}.
\texttt{MXFP} is similar to current group-wise quantization, but the difference is the shared scale is an 8-bit \texttt{float} only contains an exponent field.

\subsection{KV Cache Optimization}
Several techniques have been proposed to optimize the KV cache, including PagedAttention~\cite{kwon2023efficient}, MQA~\cite{kwon2023efficient}, GQA~\cite{ainslie2023gqa}, quantization~\cite{kang2024gear,liu2024kivi,hooper2024kvquant,zhao2023atom}, and sparsity~\cite{zhang_h_2o_2023,li_snapkv_2024,adnan_keyformer_2024,zhang_q-hitter_2024,sun_triforce_2024}.
PagedAttention is a memory management technique designed to reduce KV cache fragmentation.
Several works~\cite{kwon2023efficient,guo2024gmlake,xu2024vtensor} focus on GPU memory management to minimize the fragmentation.
In GQA and MQA, multiple query heads share a single key and value head, reducing the KV cache memory footprint.
Compression techniques like quantization and sparsity reduce the size or bit width of the KV cache.
This paper focuses on quantization, which can also be combined with other techniques to enhance memory efficiency further.
\section{Motivation}
\label{sec:motivation}

In this section, we first demonstrate that the emerging paradigm of group quantization demands a high level of adaptivity, which current adaptive methods lack.
We then discuss how adapting these methods to group quantization could compromise their efficiency.
Given that LLMs generate KV caches during runtime, real-time quantization capability is crucial.
These challenges lead to our proposal of a mathematical adaptive numerical type (\texttt{MANT}), which we will detail later.

\begin{figure}[t]
    \centering
    \begin{minipage}[t]{0.48\columnwidth}
      \centering
      \includegraphics[width=\columnwidth]{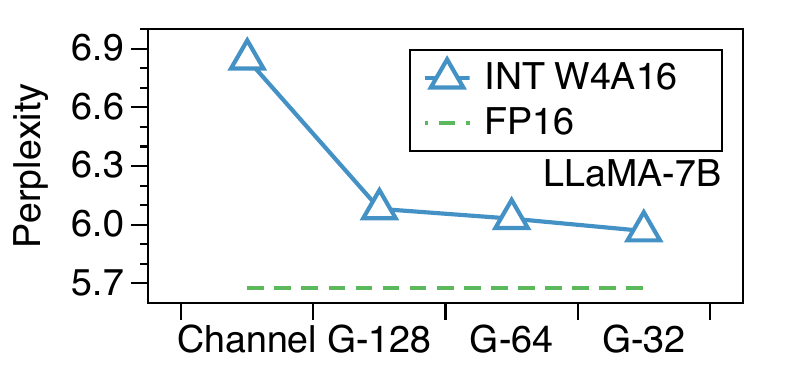}
      \caption{LLM accuracy with different quantization granularities. We report the perplexity (PPL) metric (lower is better).}\label{fig:moti_group_ppl} 
    \end{minipage}
    \hspace{2pt}
    \begin{minipage}[t]{0.48\columnwidth}
      \centering
      \includegraphics[width=\columnwidth]{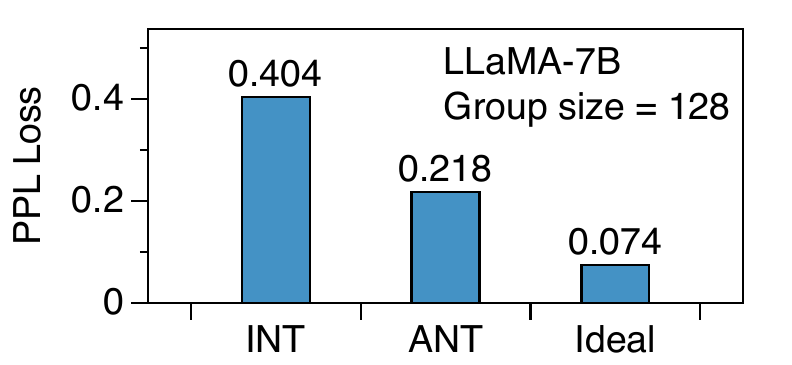}
      \caption{Accuracy loss for \texttt{INT}, \texttt{ANT}, and Ideal (clustering algorithm K-Means) adaptive methods in group quantization. }\label{fig:moti_ppl} 
    \end{minipage}
\end{figure}

\subsection{Group Quantization Accuracy Analysis}
\label{sec:acc_analysis}

In this subsection, we begin by comparing the accuracy of traditional channel-wise quantization with group-wise quantization~\cite{shao2024omniquant,zhao2023atom,liu2024kivi,sheng2023flexgen,lin2023awq,zhao2023atom}, establishing the baseline for group-wise quantization in this study.
We then delve into the use of various adaptive data types in group quantization, emphasizing the necessity for full adaptivity.

\Fig{fig:moti_group_ppl} illustrates the perplexity when quantizing the LLaMA-7B model~\cite{touvron2023llama} with various granularities using the \texttt{INT4}-based symmetric quantization.
Channel-wise quantization significantly worsens the perplexity of the examined LLM, increasing it from 5.68 to 6.85.
Conversely, group-wise quantization mitigates this loss in perplexity with a group size of 128, corresponding to an average of 4.125 bits per element (16-bit scaling factor).
Additionally, we observe that a smaller group size of 32 offers only a slight improvement in perplexity, but the scaling factor overhead increases by $4\times$.

Given this analysis, we adopt a group size of 128 as our standard configuration for the remainder of this section.
Previous research indicates that the \texttt{INT} data type is not optimal for accuracy since tensors or channels exhibit varied distributions, leading to the proposal of various adaptive data types~\cite{guo2022ant, guo2023olive, zadeh2020gobo, zadeh2022mokey}.
We evaluate their efficacy in the context of group quantization, which falls into two main categories: data-type-based and clustering-based.

\textbf{Data-type-based adaptive methods} select data types from discrete sets based on tensor data distribution.
ANT~\cite{guo2022ant} is a representative example of the data-type-based method.
ANT packages several different data types for selection, including \texttt{INT} for the uniform distribution, \texttt{PoT} (Power of Two) for the Laplace distribution, and \texttt{flint} for the Gaussian distribution.

\textbf{Clustering-based adaptive methods} utilize clustering algorithms to generate centroids that align with the data distribution and provide considerable adaptivity. 
Mokey~\cite{zadeh2022mokey} and GOBO~\cite{zadeh2020gobo} exemplify this approach, though they focus on tensor- or channel-wise quantization. In our study, we adapt them to group quantization through per-group clustering.

\Fig{fig:moti_ppl} compares the accuracy of the methods described above for the LLaMA-7B model under 4-bit group-wise quantization. 
The group-wise \texttt{ANT} method outperforms the \texttt{INT} type by dynamically selecting from three data types to better match the value distribution, resulting in reduced perplexity (PPL) loss. 
Moreover, per-group clustering adjusts more effectively to the value distribution of each group, establishing itself as the accuracy-optimal and ideal adaptive method. 
This approach achieves nearly lossless 4-bit quantization, equivalent to 16 centroids per group. 
However, this ideal scenario is impractical due to the significant overhead associated with storing per-group centroids, effectively rendering it a 6-bit quantization.

\begin{figure}[t] 
    \centering 
    \includegraphics[width=1.0\linewidth]{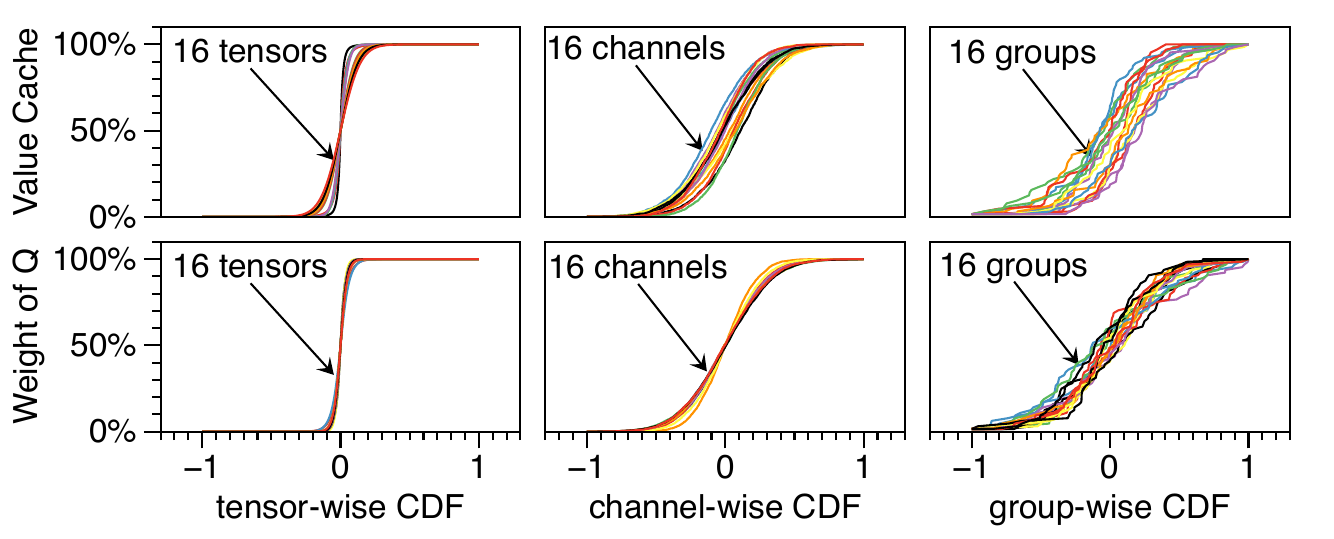}  
    \caption{The cumulative distribution function (CDF) of the tensor, channel, and group, respectively. The tensor data were taken from layers 8 to 23, while the 16 channel and group data were sampled from one tensor with specific strides.}\label{fig:moti_dist} 
\end{figure}

To illustrate the group-wise diversity in data distribution, we sampled the weights of the Q and V tensors in LLaMA-7B model. 
We normalized all sampled data to their absolute maximum values, which ranged from -1 to 1. \Fig{fig:moti_dist} displays the cumulative distribution function (CDF) for the tensor, channel, and group levels, respectively. 
We observed that the diversity at the group level is significantly higher than at the tensor level. 
In simpler terms, while different tensors exhibit similar distributions, groups can have markedly different distributions. This finding underscores the necessity for full adaptivity in group quantization to fully realize its potential.
\paragraph{Takeaway 1.} The group quantization is an emerging paradigm to accelerate LLMs, and the significant group-level diversity requires a high level of adaptivity to fully unleash its potential.

\subsection{Group Quantization Efficiency Analysis}
\label{subsec:efficiency}

In this subsection, we provide a detailed efficiency analysis for the above adaptive quantization methods.
In \Tbl{intro:dtype}, we compare OliVe~\cite{guo2023olive}, ANT~\cite{guo2022ant}, GOBO~\cite{zadeh2020gobo}, and Mokey~\cite{zadeh2022mokey} with \texttt{INT} regarding the efficiency of computation, encoding, and decoding. 
In this paper, we use the term encoding (decoding) interchangeably with quantization (dequantization).

Data-type-based adaptive methods such as ANT~\cite{guo2022ant} and Olive~\cite{guo2023olive} achieve computational efficiency comparable to \texttt{INT}. 
Both utilize specialized decoders that decode these data types prior to computation, resulting in high decoding efficiency. 
However, as previously demonstrated, these methods suffer from limited adaptivity in the group quantization paradigm. 
A straightforward approach to enhance adaptivity is to expand their set of data types. 
However, incorporating new data types necessitates additional decoders, escalating hardware design costs. 
Additionally, compatibility issues between new and existing data types may reduce computational efficiency. 
For instance, the \texttt{NF4} data type~\cite{dettmers2023qlora} requires an FP16 MAC unit, which is incompatible with existing \texttt{ANT} data types.

\paragraph{Takeaway 2.} Enhancing the data-type-based adaptive method for group quantization is challenging and requires a careful balance for the computation and decoding efficiency.

Clustering-based adaptive methods like GOBO~\cite{zadeh2020gobo} and Mokey~\cite{zadeh2022mokey} can sufficiently adapt to various distributions at the group level. 
However, they require codebooks for quantization and dequantization, leading to high adaptivity at the expense of encoding and computational efficiency. 
For instance, a 16-entry codebook with 8 bits per entry requires 128 bits per group, creating an inevitable trade-off between adaptivity and memory overhead. GOBO~\cite{zadeh2020gobo} employs the K-means algorithm to quantize weights and requires dequantization to \texttt{FP16} using a codebook lookup table before computation, resulting in high adaptivity but low computational efficiency. 
Conversely, Mokey~\cite{zadeh2022mokey} enhances the computation of clustering-based methods by using indices for centroid values via approximate calculations, though matrix multiplication still relies on floating-point units, increasing overhead compared to integer units. 
Furthermore, Mokey creates one \texttt{golden dictionary} for all activations and weights, akin to using a single data type in quantization, thus reducing adaptivity.

\paragraph{Takeaway 3.} Deploying the clustering-based adaptive methods under group quantization is challenging owing to the low encoding and computation efficiency.

\begin{table}[t]
    \centering
    \small
    \renewcommand{\arraystretch}{1.2}
    \caption[]{Features of DNN accelerators with adaptive and flexible data types are summarized. Here, `Effi.' stands for efficiency, `Med.' for medium, and `LUT' for lookup table.}
  
    \resizebox{1.0\columnwidth}{!}{
      \begin{tabular}{c|cc|ccc|cc|c}
        \Xhline{1.2pt}
        \multirow{2}{*}{Architecture} & \multicolumn{2}{c|}{Encode} & \multicolumn{3}{c|}{Computation} & \multicolumn{2}{c|}{Decode} & \multirow{2}{*}{Adaptivity} \\ \cline{2-8}
        & Method & Effi. & Method & Bit & Effi. & Method & Effi. \\
        \Xhline{1.2pt}
        \texttt{INT} & Round & High & INT & 4 \& 8 & High & Calculation & High & Low \\ 
        OliVe~\cite{guo2023olive} & Search & Med. & INT & 4 \& 8 & High & Decoder & High & Med. \\ 
        ANT~\cite{guo2022ant} & Search & Med. & INT & 4 \& 8 & High & Decoder & High & Med. \\ 
        Mokey~\cite{zadeh2022mokey} & Cluster & Med. & Float & 4 \& 8 & Med. & Calculation & Med. & Low \\ 
        GOBO~\cite{zadeh2020gobo} & Cluster & Low & Float & 16 & Low & LUT & Med. & High \\ 
        \hline
        \multirow{2}{*}{\proj}  & Search  & Med.  & \multirow{2}{*}{INT} & \multirow{2}{*}{4 \& 8} & \multirow{2}{*}{High} & \multirow{2}{*}{Calculation} & \multirow{2}{*}{High} & \multirow{2}{*}{High} \\ \cline{2-3}
        &  Map &  High &  &&&\\ 
        \Xhline{1.2pt}
    \end{tabular}
    }
    \vspace*{0.1cm}
    \label{intro:dtype}
    \vspace*{-0.2cm}
  \end{table}

\subsection{Support for Real-time Quantization}
\label{sec:moti_kvcache}

The above group-wise diversity presents a challenge for both weights and KV cache.
In addition, KV cache faces challenges in real-time group-wise quantization because the KV cache is generated dynamically during LLM inference.

\begin{figure}[t] 
  \centering 
  \includegraphics[width=0.9\linewidth]{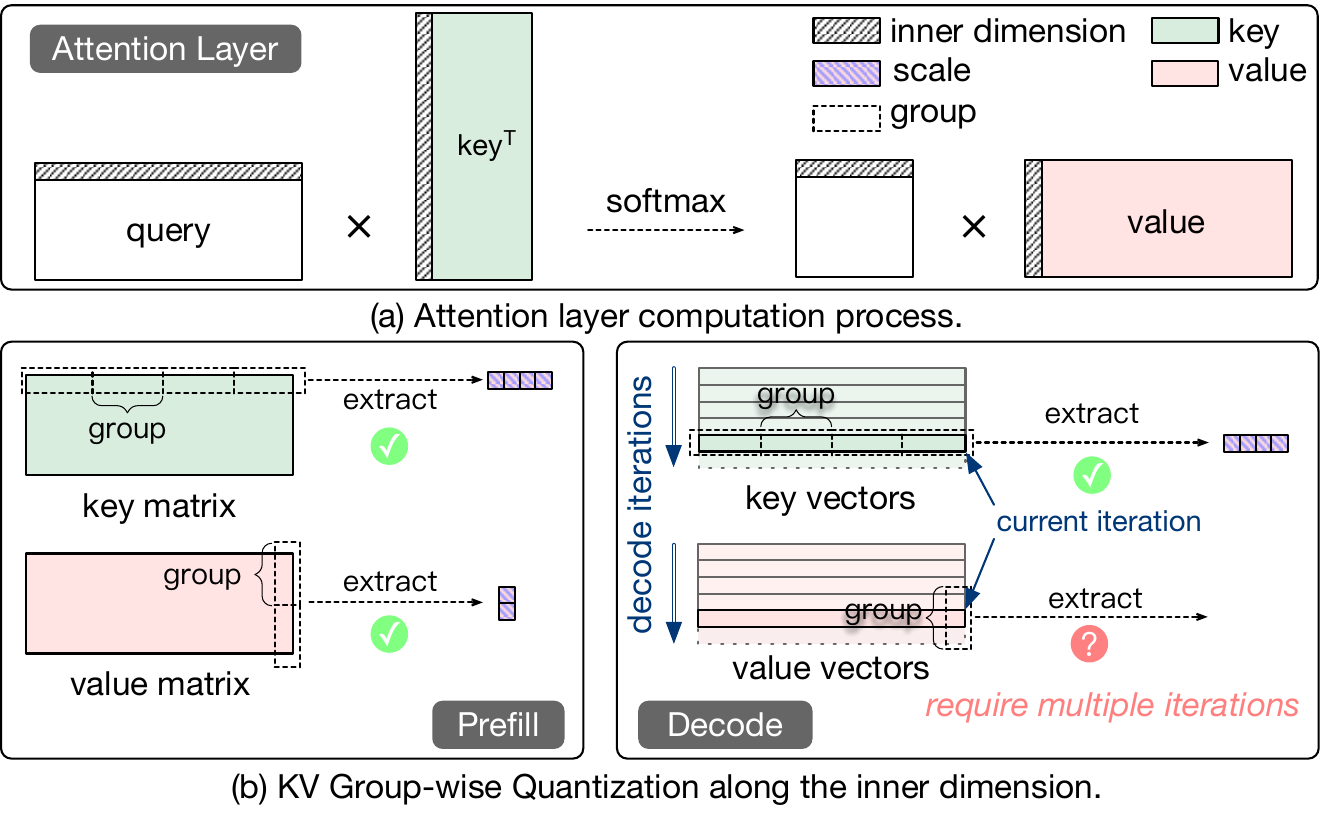}  
  \caption{\small Comparison of group-wise K and V cache quantization. They have different inner dimensions due to the transposition of K (key).}

  \label{fig:kv_process}
\end{figure}

To facilitate low-precision computation in group-wise quantization, it is necessary to quantize K and V along the inner dimension. 
This requirement stems from the support for matrix inner product operations in most GPUs and TPUs. 
During these operations, the group-wise scaling factor can be extracted from the multiply-accumulate process. 
\Fig{fig:kv_process} depicts the computation process of K and V during the decode stage. We define the dimension used for matrix inner product operations as the inner dimension. 
The inner dimensions of the K and V caches differ; the K cache requires a transpose operation, whereas the V cache does not, complicating the situation.

In the prefill stage, K and V can easily compute the scaling factor for each group. 
During the decode stage, the newly generated K vector is concatenated along the inner dimension of the K cache, enabling immediate quantization. 
However, the newly generated V vector is associated with different groups, with only one element per group produced per iteration. This process prevents the scaling factor for the entire group from being obtained in a single iteration, posing a significant challenge for the real-time quantization of the V cache.

Given those challenges, we propose \proj with a mathematical encoding format that can fuse with integer computation and enhance the decoding efficiency.
In addition, this encoding format provides sufficient adaptivity for group-wise quantization.
Regarding the challenge in KV cache, \proj employs a real-time quantization engine that ensures efficient encoding and decoding for KV cache.
By addressing these challenges, \proj enables efficient low-bit group-wise quantization.

\section{Mathematically Adaptive Numerical Type}
\label{sec:encode}


We propose \proj (mathematically adaptive numerical type), a directly computable and adaptive data representation that achieves efficient encoding/decoding under the group-wise quantization framework.
We first present the formal mapping of \proj{}-encoded \texttt{INT} values in \Sec{sec:encode_mapping}.
We then describe its efficient encoding process in \Sec{sec:encode_encode} and how it eliminates the necessity for a data type-specific decoder in \Sec{sec:encode_decode}.

\subsection{Mapping Representation}
\label{sec:encode_mapping}

Two key considerations drive the design of our mapping representation.
The first consideration is the ability to accommodate the diverse distributions of group-wise data.
The second consideration is the computation efficiency, suggesting a direct computation approach using the \texttt{INT} value.
Based on the two considerations, we propose a mapping representation defined by the following equation:
\begin{equation}	
    Value_{grid} = \pm (a \times |\texttt{INT}| + 2^{|\texttt{INT}|})
    \label{eqn:map_represent}
\end{equation}

In Equation~\eqref{eqn:map_represent}, the $Value_{grid}$ is the quantization grid, $a$ is a group-wise constant, and \texttt{INT} is a sign-magnitude representation of \texttt{INT4} within the context of symmetric quantization, which covers the range from [-7, 7].
We take \texttt{INT4} as an example to explain our design since the computation of \texttt{INT4} is easy to support and \texttt{INT4} is memory-aligned.
The quantization grid, denoted as $Value_{grid}$, is constructed as follows: $\{\pm (a \times 0 + 2^{0}), \pm (a \times 1 + 2^1),..., \pm (a \times 7 + 2^7) \}$.

\begin{figure}[t] 
    \centering 
    \includegraphics[width=0.9\linewidth]{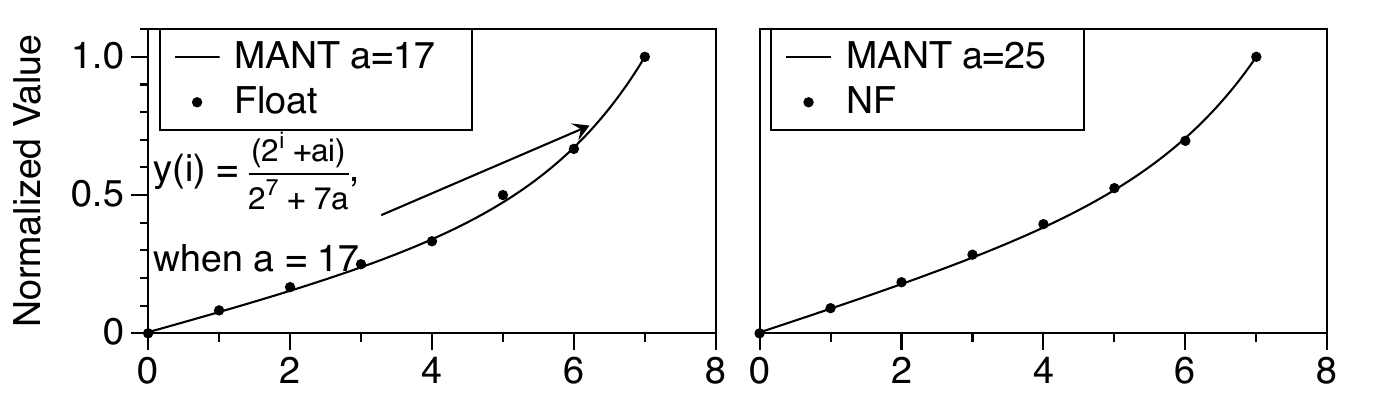}  
    \caption{Using different $a$ in \proj{} for data type approximation.}
    \label{fig:encode_function}
\end{figure}

Our core idea is to approximate each data type by changing the coefficient $a$.
The following equation defines the positive value function of different data types.
\begin{equation}	
\label{eq:fucntion}
    \begin{array}{c}
    y_{\text{INT}}(i) = i, y_{\text{PoT}}(i) = 2^i, y(i)_{\text{MANT}} = ai + 2^i, i \in [0, 7] \\ 
%
    y_{\text{NF}}(i) = \Phi^{-1}\left(\frac{i \times (1-\epsilon) \times 0.5 }{7} + 0.5\right), \quad i \in [0, 7]
\end{array}
\end{equation}
Here, integer $i$ ranges from 0 to 7, and $\Phi^{-1}$ represents the probit function, the inverse of the cumulative Gaussian distribution.
The small $\epsilon$ prevents $\Phi^{-1}$ from reaching $\infty$.
To represent a given datatype in \proj{}, we only need to find a proper coefficient $a$ that minimizes the approximation error.
For example, to represent $y_{\text{INT}}(i)$, the goal is to minimize the target function $\mathop{argmin_a} (|\frac{i}{7} - \frac{ai + 2^i}{7a+2^7}|)$, where the division by $7$ and $7a+2^7$ serves to normalize and match the maximum values of the distributions.
\Fig{fig:encode_function} shows the values of $a$ to represent \texttt{Float} and \texttt{NF}.

\begin{figure}[t] 
    \centering 
    \includegraphics[width=1\linewidth]{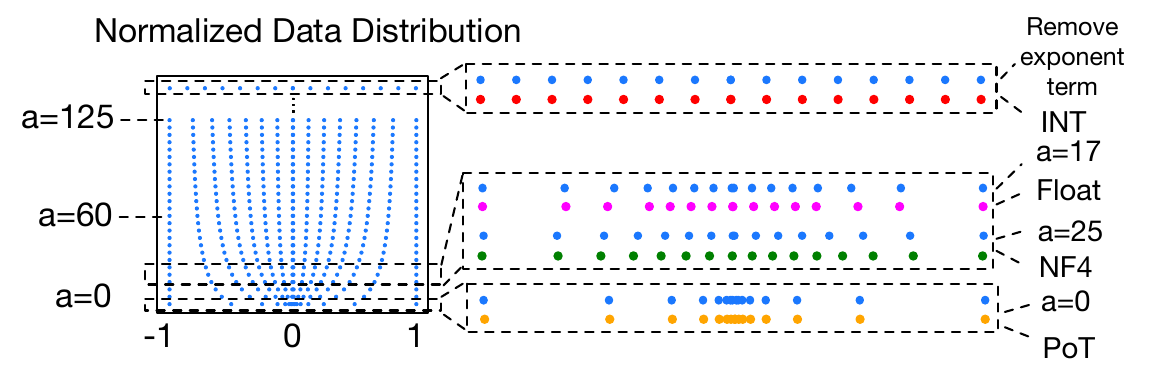}  
    \caption{The diversity of data representation with coefficient $a$. The 4-bit data types have 16 points. Left part of figure shows how the distribution varies with the increase of a. Right part of figure shows that when setting a to a different value, it can nearly match the distribution of several data types, including \texttt{float}, \texttt{NF}, \texttt{PoT}, and \texttt{INT}.}
    \label{fig:distribution}
\end{figure}

\Fig{fig:distribution} shows the normalized data distribution across various coefficient $a$ values.
Modifying the coefficient $a$ leads to a smooth change in the data distribution, allowing for a versatile mapping representation that accommodates a wide range of data types.
For instance, setting the coefficient $a$ to 0 transforms the mapping representation to $Value = \pm (2^{|\texttt{INT}|})$, it makes \proj exactly match the data type \texttt{PoT}~\cite{miyashita2016convolutional, zhou2017incremental}.
Besides, \proj can approximate the distribution of \texttt{float} and \texttt{NormalFloat (\texttt{NF})}~\cite{dettmers2023qlora} when setting coefficient $a$ to 17 and 25, respectively.

It is worth noting that the role of coefficient $a$ is to change the data distribution. 
Meanwhile, coefficient $a$ can be integrated into the computing process of dequantization with low computation overhead, which we detail later. 
Our experiment finds that the variation of the data distribution becomes marginal when the coefficient $a$ exceeds 128. 
As such, we constrain the data range of $a$ within 128, allowing 8-bit encoding for $a$.

\subsection{Encode}
\label{sec:encode_encode} 

Based on the mapping of \proj, we describe its encoding process, assuming that activations are quantized to 8 bits and weights to 4 bits with a determined coefficient $a$ from original FP16 values.
In the group-wise quantization, each group stores the metadata that includes both the scaling factor and the coefficient $a$.
The quantization process is defined as:
\begin{equation}	
    \begin{array}{c}
    s_X = \frac{max(|X_{FP16}|)}{max(INT8)},  s_W = \frac{max(|W_{FP16}|)}{max(W_{grid})}    \vspace*{0.2cm} \\
    X_{INT8} = \lfloor \frac{X_{FP16}}{s_{X}} \rceil, W_{INT4} = argmin(\frac{W_{FP16}}{s_{W}} - W_{grid})
    \end{array}
    \label{eqn:quantize_process}
\end{equation}

$X$ represents the activation and $W$ represents the weight.
$s_X$ is the scaling factor of activations and $s_W$ is the scaling factor of weights, while $W_{grid}$ is the quantization grid with specific $a$.
\Fig{fig:encode_process} shows an example of \proj with $a=17$.
The rounding and encoding process of weights in \Fig{fig:encode_process} corresponds to $argmin$ operation in the Equation~\eqref{eqn:quantize_process}.
The encoding process is expensive since it necessitates finding the nearest point within a non-uniform grid.
Nonetheless, the weights encoding process can be done offline, avoiding the runtime overhead.

\subsection{Decode \& Compute Fusion}
\label{sec:encode_decode}

\begin{figure}[t] 
    \centering 
    \includegraphics[width=0.9\linewidth]{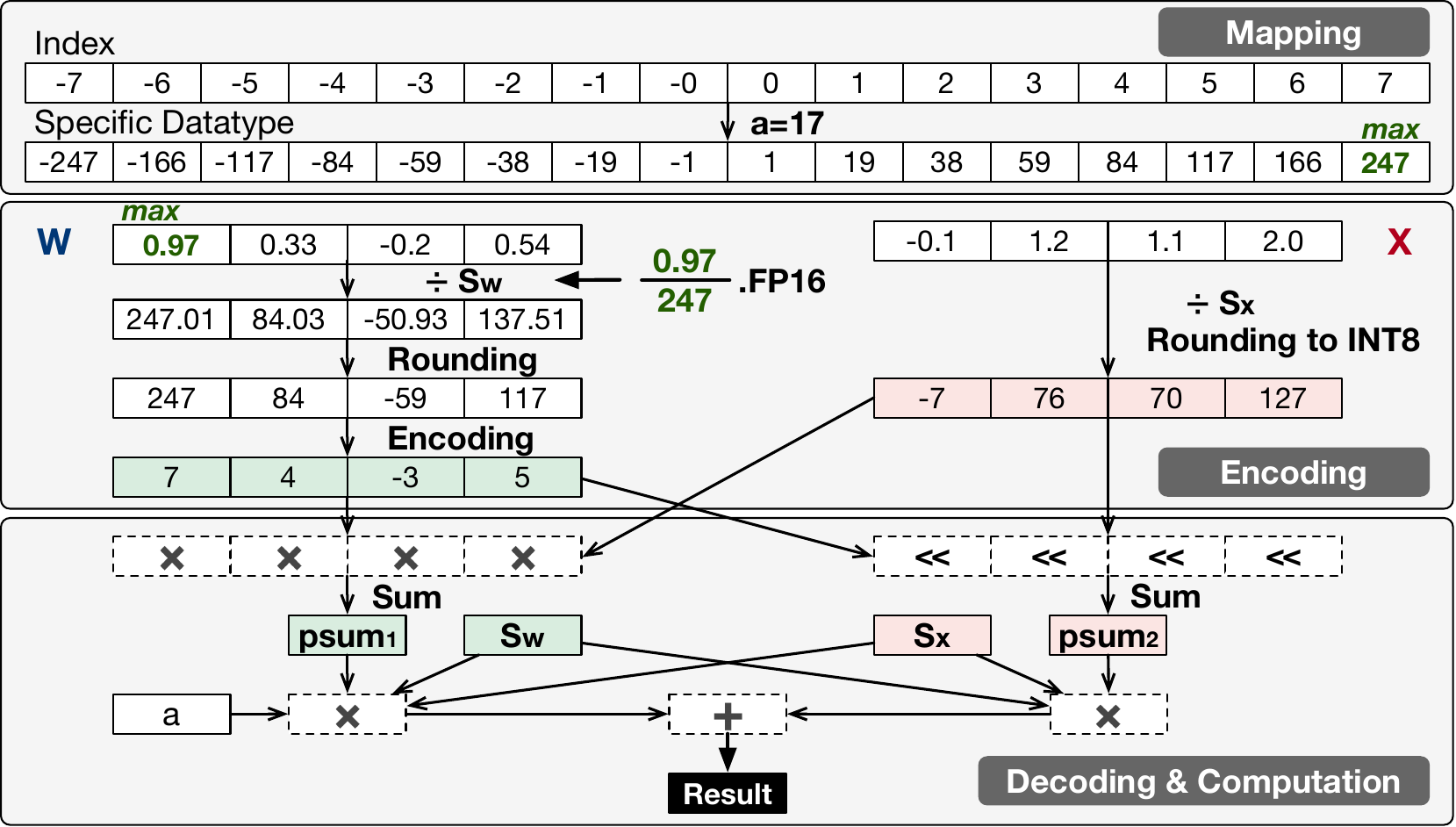}  
    \caption{The encoding and decoding process of \proj.} 
    \label{fig:encode_process}
\end{figure}

We use the example in \Fig{fig:encode_process} to illustrate the fused decoding and computing process.
The primary advantage of \proj lies in its ability to perform computations efficiently in low-bit formats, with the decoding process integrated into matrix multiplication.
In our approach, the activations and weights are quantized along their cumulative dimensions, allowing to decouple the computation of the scaling factors $s_{X}$ and $s_{W}$ from multiplication and addition.
The combined decoding and computing process is described by the equation below:
\begin{equation}
    \begin{array}{ll}
    &\phantom{}\hat{X}_{FP16} \times \hat{W}_{FP16} \\
    &\phantom{}= (X_{INT8} \times s_{X}) \times (W_{grid} \times s_{W}) \\
    &\phantom{}= [X_{INT8} \times W_{grid}] \times s_{X} s_{W} \\
    &\phantom{}= [X_{INT8} \times (a \times W_{INT4} + 2^{W_{INT4}})] \times s_{X} s_{W} \\
    &\phantom{}= \underbrace{[X_{INT8} \times W_{INT4}]}_{\text{$psum_1$}} \times a \cdot s_{X} s_{W} + \underbrace{[X_{INT8} \times 2^{W_{INT4}}]}_{\text{$psum_2$}} \times s_{X} s_{W} 
    \end{array}
    \label{eqn:decode}
\end{equation}

$\hat{X}_{FP16}$ and $\hat{W}_{FP16}$ are dequantized activations and weights, respectively.
In Equation~\eqref{eqn:decode}, $W_{grid}$ is derived from Equation~\eqref{eqn:map_represent}.
Thus, the computation can be divided into integer multiplication and integer shift operations.
Moreover, it facilitates computations in a mixed-precision mode, using 8-bit activations and 4-bit weights, eliminating the need to dequantize low-bit weights before computation.    
To simplify the discussion, we omit the details of handling the sign bit in Equation~\eqref{eqn:decode}, which can be efficiently processed in hardware.

\section{\proj Quantization Framework}
\label{sec:dse}

This section introduces the \proj quantization framework, focusing on the quantization methods for weight, activation, and KV cache.
Initially, we discuss the weight quantization with \mbox{\proj} in \mbox{\Sec{sec:weight_quant}}.
Subsequently, we illustrate why activation is quantized with \texttt{INT8} in \Sec{sec:dse_act}.
Finally, we introduce a real-time \proj quantization mechanism to tackle the challenge brought by the dynamic KV cache and customize quantization strategies for K and V cache, respectively, in \Sec{sec:dse_kv}.


\subsection{Weight Quantization}
\label{sec:weight_quant}

Weight is encoded offline to select the most suitable coefficient $a$ for each group.
Within the \proj framework, a calibration dataset~\cite{gao2020pile} is used to identify the coefficient $a$ that minimizes the mean square error (MSE) of the output.
The following equation describes the optimization objective:
\begin{equation}	
    a = \mathop{argmin}\limits_{a} ||X \hat{W}_{a} - X W||^2_{2}
    \label{eqn:min_mse}
\end{equation}
$\hat{W}_{a}$ represents the weight that is first quantized and subsequently dequantized using a specific coefficient $a$.
Broadening the range of data types in the search space typically enhances accuracy.
Nevertheless, slight modifications to $a$, increasing or decreasing it by one, only slightly alters the data distribution.
Consequently, we selected 16 data types for weight quantization, including the set $\{0, 5, 10, 17, 20, 30, 40, 50, 60, 70, 80, 90, 100, 110, 120\}$ and an additional \texttt{INT} option.

\subsection{Activation Quantization}
\label{sec:dse_act}

For the activation quantization, we follow the common practice that quantizes them using the group-wise \texttt{INT8} for near-lossless accuracy and efficient hardware implementation~\cite{frantar2023gptq,lin2023awq, kim2023squeezellm}.
Meanwhile, unlike weight and KV cache, activations are temporary variables.
They are dynamically generated, quantized, and consumed, so they cannot be reused in the latter inference iteration.
Finally, the weights and KV caches already consume most of memory in the LLM inference so that activations occupy only a minor fraction of the memory ($<$5\%~\cite{yuan2024llm}).
As such, further reducing the activation bit width only leads to marginal improvement.

This choice of using \texttt{INT8} to activation quantization also allows the computation between the weight that quantized using our \proj{} and the activation that quantized using \texttt{INT8} to exploit the low-bit computation units, as detailed in \mbox{\Sec{sec:encode_decode}}.
\proj further leverages a streaming comparator unit to determine the maximum value of activations, facilitating the activation quantization.

Real-time quantizing activation requires two operations: derive the maximum value of a group to calculate scaling factor and map the \texttt{FP16} value to \texttt{INT8}.
\proj hides the latency of searching the maximum value with hardware design fusing this operation into the streaming of systolic array output, as detailed in \Sec{sec:real_time_engine}.

\subsection{KV Cache Quantization}
\label{sec:dse_kv}

\paragraph{The Quantization Dimension.}
We quantize K and V along the accumulation dimension, as we discuss in \Fig{fig:kv_process}.
Some prior algorithmic works~\cite{liu2024kivi,hooper2024kvquant} do use a different quantization direction for V cache, which we believe is orthogonal to our work. 
First, MANT can also work with the channel direction of V cache quantization, which may sacrifice the computation efficiency of all quantization methods, including ANT. 
Second, those prior works are based on channel-level quantization or extremely low bit, while the impact of directions for the 4-bit group-level quantization is much smaller, as we show later. 
Finally, emerging incoherent processing algorithms~\cite{ashkboos2024quarot,tseng2024quipbetterllmquantization,xiao2023smoothquant} (where SmoothQuant~\cite{xiao2023smoothquant} is a special case) are very promising to further mitigate this gap.

\paragraph{Select \proj through Variance.}
In KV cache quantization, real-time data type selection is required.
Although the searching method based on MSE achieves less accuracy loss, it requires performing quantization to each data type for MSE searching, which is intolerable in a real-time scenario.
Thus, we develop a mapping mechanism based on the data characteristics like variance, which can be derived in a streaming way.

Since variance can reflect the distribution of a data group and \proj with different $a$ has a different variance, \proj determines the coefficient $a$ using a variance.
Then, calculate the variance:
\begin{equation}
    \sigma^2 = \frac{1}{n} \sum_{i=1}^{n} x_i^2 - \left( \frac{1}{n} \sum_{i=1}^{n} x_i \right)^2
    \label{eqn:variance}
\end{equation}

In \proj, a larger $a$ corresponds to higher variance, with each value of $a$ associated with a specific variance range.
We first sample the K and V tensors through a calibration dataset~\cite{gao2020pile}, and select $a$ for each group to minimize quantization error.
Next, we calculate the variance of the groups with different $a$ to decide the appropriate range.
The elements within each group are first normalized, ensuring that the absolute maximum value in the group is scaled to 1. 
Following this normalization step, the variance of the normalized elements is then computed.
For example, when $a=35$, the variance is 0.104; when $a=45$, the variance is 0.118.
We define the range for $a=40$ as [0.104, 0.118].
If the variance of normalized data falls within a specific range of $a$, it will be quantized with that $a$ with \proj.
We fuse the computation flow of variance with matrix multiplication to hide latency, which we will detail in \Sec{sec:real_time_engine}.

\paragraph{Prefill Stage.}
During the prefill stage, the input comprises a sequence, so the K and V are both matrices, where the sequence length typically exceeds the group size.
Thus, both the K cache and V cache can obtain the data needed to calculate variance.
By selecting the appropriate $a$ based on the variance, the K cache and V cache can be quantized to 4-bit \proj.

\paragraph{Decode Stage.}
In the decode stage, since inputs are vectors, generated K and V are vectors.
Thus, the K cache can obtain all data of the group in a single iteration to execute real-time quantization, similar to activation in decode stage.
The difference is that each group needs to calculate the partial sum of $x_i$ and $x_i^2$ as well as the maximum value since K cache needs to be quantized to 4-bit \proj.
However, when it comes to V cache, a new challenge arises as each iteration only generates one element of a group.

\begin{figure}[t] 
    \centering 
    \includegraphics[width=0.9\linewidth]{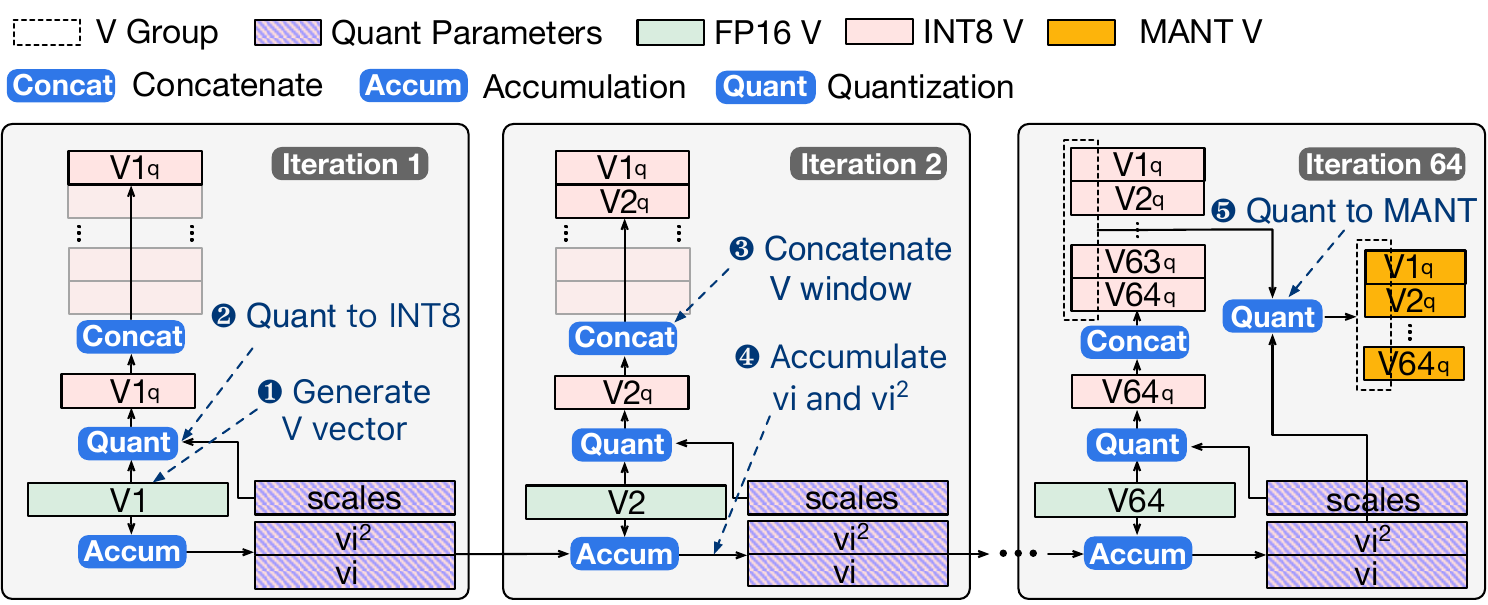}  
    \caption{The update process of V cache with a group size of 64. Each iteration generates a new $V$ vector, quantizes it to INT8, and accumulates the parameters for \proj quantization. In the 64th iteration, the V cache is quantized to 4-bit \proj.}
    \label{fig:v_update}
\end{figure}

To address this, we propose a two-phase quantization scheme for the V cache, as shown in \Fig{fig:v_update}.
We define every $G$ iterations in the decode phase as a process window for V cache, where $G$ is the group size. 
In the first phase, newly generated V vector is quantized to \texttt{INT8} with channel-wise scaling factors derived from prefill stage, denoted as `scales' in \Fig{fig:v_update}. 
Meanwhile, we update the maximum value and partial sum of $v_i$ and $v_i^2$, denoting the parameter in a group as $v_i$.
This operation persists until the process window is full.

The second phase involves quantizing the 8-bit V cache to 4-bit \proj.
When the process window is full, we calculate the variance with partial sum of $v_i$ and $v_i^2$ by ~\eqref{eqn:variance}.
Then, the coefficient $a$ is decided similarly to the prefill phase, and the stacked \texttt{INT8} V cache is quantized to 4-bit \proj.

This two-phase quantization scheme effectively quantizes all but the latest V vectors in the processing window to 4 bits.
The process window is similar to the residual group used in KIVI~\cite{liu2024kivi}.
The overhead of \texttt{INT8} operation of V cache in processing window is marginal and tolerable.
Besides, it helps improve the quality of newly generated tokens, since some studies have shown that the latest tokens are more important~\cite{duanmu2024skvqslidingwindowkeyvalue}.

\section{\proj Microarchitecture}
\label{sec:architecture}


This section outlines the microarchitectural details of \proj{} that enable efficient quantization for LLM inference. Our hardware design supports the fusion of \proj{} decode and computation, as detailed in \Sec{sec:encode}. Additionally, it integrates microarchitectural components that facilitate real-time quantization of activations and the KV cache, as described in \Sec{sec:dse}.

\subsection{Accelerator Overview}

We first explain the rationale for the architectural extension. 
Initially, the computing paradigm of \proj{} introduces an additional term in matrix operations, which the current Processing Element (PE) does not directly support. 
Secondly, we need an on-chip quantization engine to minimize real-time quantization delays for activation and KV cache. 
Finally, our quantization framework uses varying bit widths for weights, activations, and the KV cache, necessitating mixed-precision computations.

\begin{figure}[t] 
    \centering 
    \includegraphics[width=0.9\linewidth]{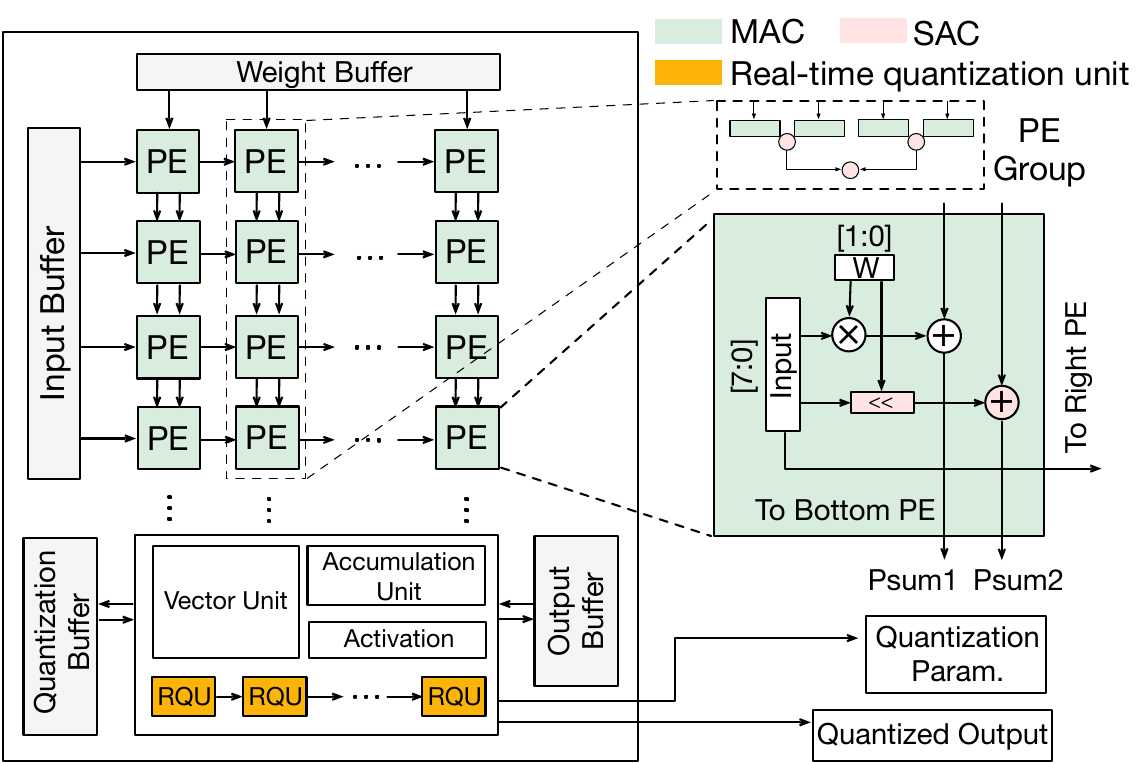}  
    \caption{Accelerator overview. MAC and SAC are multiply-accumulate and shift-accumulate units, respectively.}
    \label{fig:arch_overview}
\end{figure}

\Fig{fig:arch_overview} provides an architectural overview that adopts a weight-stationary systolic array with only slight modifications.
The array consists of $32 \times 32$ units, each capable of handling 8-bit integers, referred to in this section as Processing Elements Group (PEG).
The real-time update engine incorporates a lightweight comparator unit to identify maximum values.
During dequantization, output values are multiplied by a scaling factor and processed through this comparator to determine maximum values for use in the quantization units.

\subsection{\proj PE Unit}
\label{sec:pe_unit}
\paragraph{Tile Architecture.}
We begin by describing the tile architecture of \proj, which is based on a weight-stationary systolic array~\cite{jouppi2023tpu}. Inputs to the systolic array are fed from the left, and outputs exit from the bottom. 
The array's configuration varies with the type of operation: a 32 $\times$ 32 array for \texttt{INT8} $\times$ \texttt{INT8} operations, a 64 $\times$ 32 array for \texttt{INT8} $\times$ \texttt{INT4} operations, and a 128 $\times$ 32 array for \texttt{INT8} $\times$ \texttt{INT2} operations. 
This variation in configuration is due to the architecture's support for mixed-precision operations.
Specifically, the bit width of weight can be configured to 2, 4, or 8 bits.

Our discussion will focus on \texttt{INT8} and \texttt{INT4} operations, where the weight tile dimension is $(64, 32)$, the input tile dimension is $(m, 64)$, and the output tile dimension is $(m, 32)$. In this configuration, each column of PEs represents the cumulative dimension, with elements in a column sharing both a scaling factor and the coefficient $a$ of \proj.

Each PE in our architecture includes both multiply-accumulate (MAC) and shift-accumulate (SAC) components. The MAC components manage the first partial sum ($psum_1$) defined in Equation~\eqref{eqn:decode}, while the SAC handles the second partial sum ($psum_2$). Executing $psum_1 \times a$ within the PE facilitates the addition of $psum_1$ and $psum_2$. However, this process introduces an additional multiplication step across all PEs. To streamline computations in the systolic array, \proj directs $psum_2$ through an additional lane.

\paragraph{Mixed Precision.}
Given the varying bit-width requirements for weight, activation, and KV cache quantization, support for mixed precision is crucial, as noted in previous studies~\cite{guo2022ant,guo2023olive,song2020drq,zheng2022dota,zhou2016dorefa,micikevicius2018mixed,cai2020zeroq,cai2020rethinking,Reggiani2023mixgemm}. Our approach aligns with methodologies found in BitFusion~\cite{sharma2018bit}. As depicted in \Fig{fig:arch_overview}, each Processing Elements Group (PEG) within our system includes four Processing Elements (PEs), each capable of handling computations between \texttt{INT8} and \texttt{INT2} formats. Two PEs combined can perform an \texttt{INT8} $\times$ \texttt{INT4} operation. A single
PEG can execute either one \texttt{INT8} $\times$ \texttt{INT8} operation or four \texttt{INT8} $\times$ \texttt{INT2} operations within one cycle. This configuration efficiently supports mixed-precision computations across 2, 4, and 8-bit widths for weights and KV cache.

\subsection{Real-time Quantization Unit}
\label{sec:real_time_engine}
We describe how our architecture implements real-time quantization for activations, K cache, and V cache. 
We leverage the dataflow~\cite{guo2020balancing,jouppi2023tpu,zhou2021characterizing,chen2016eyeriss,VELTAIR,zhou2023ugrapher} of matrix multiplication to hide latency.
In this scenario, the group size is set to 64.

\begin{figure}[t] 
    \centering 
    \includegraphics[width=1\linewidth]{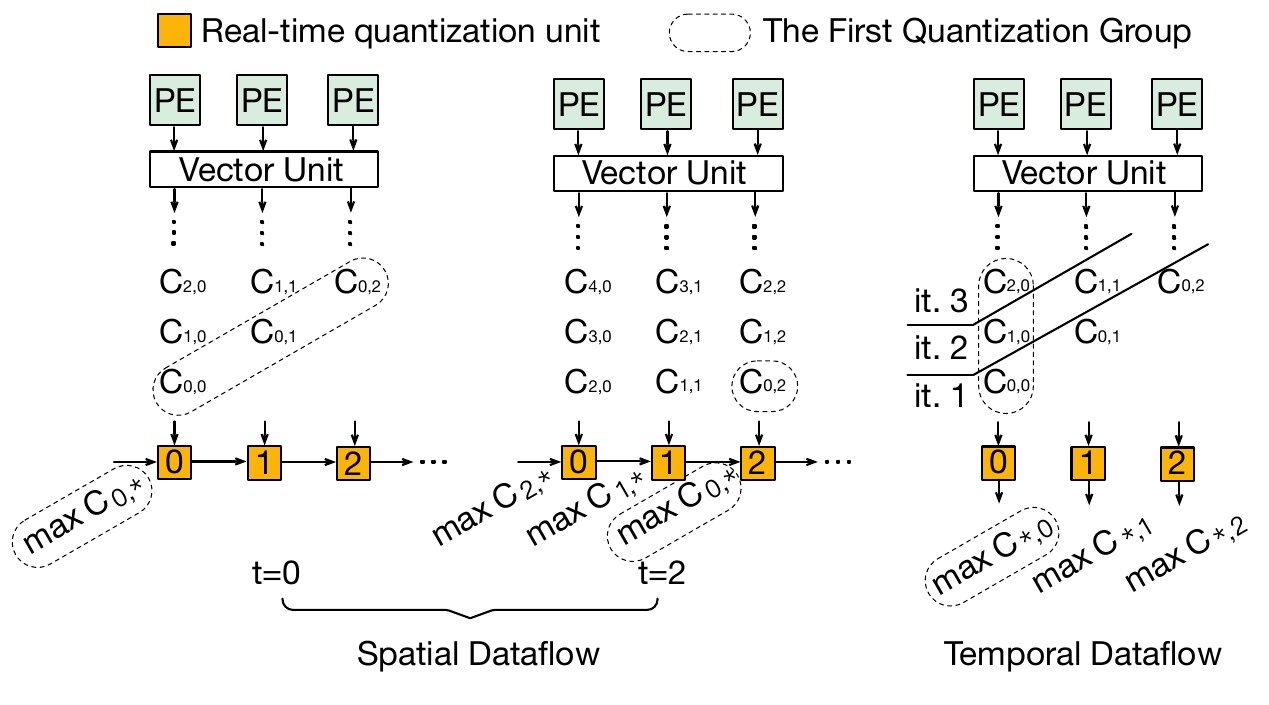}  
    \caption{Spatial and temporal dataflow mode of real-time quantization unit. ``It.1'' in this figure means the first iteration for V group during the decode stage of LLM inference.} 
    \label{fig:arch_cmp}
\end{figure}

Our real-time quantization unit (RQU) consists of two components: an \texttt{FP16} comparator and two \texttt{FP16} accumulators.
In \proj, we utilize 32 RQUs matching the size of the systolic array to determine the maximum values and calculate variance parameters for the K cache and V cache.
Besides, it can derive the maximum values for \texttt{INT} activation quantization.
The RQU supports both \textbf{spatial} and \textbf{temporal} dataflow modes. 
\Fig{fig:arch_cmp} illustrates these two modes and shows an example where RQUs are used to compute their maximum values.
Once the systolic array completes the final computation in the accumulation dimension, it captures the full results of the output tile, which are then quantized to a low-bit format. 
In the weight-stationary systolic array, the leftmost column initiates the computation, and the rightmost column begins 31 cycles later, offering a chance to pipeline the process of maximum comparison with the preceding computations.

As depicted in \Fig{fig:arch_cmp}, at $t=0$, the first RQU ($RQU_0$) uses $C_{0, 0}$ as the initial maximum value, $max(C_{0, *})$, and passes it to $RQU_1$. 
At $t=1$, $RQU_1$ compares the value of $C_{0, 1}$ with $max(C_{0, *})$ and forwards the greater value to $RQU_2$. By $t=32$, $RQU_{31}$ outputs the final maximum value $max(C_{0, *})$. 
From this point, the RQUs operate in a fully pipelined manner, with $RQU_{31}$ producing one maximum value per cycle. Given that the output tile's dimensions are $(m, 32)$ (with $m=1$ during the decode phase), identifying the maximum of 64 elements in one group requires two comparison rounds. 
Subsequently, activation can be quantized based on their maximum values.

Conversely, the V cache utilizes the \textbf{temporal} dataflow mode of RQU, as illustrated on the right side of \Fig{fig:arch_cmp}. 
In this mode, each RQU compares values from the same column of the systolic array and retains the maximum value in its register. 
During the prefill phase, each RQU captures the maximum value of every 64 elements, allowing the 64 elements from the same column to be quantized based on these maximum values.

The dataflow of calculating sum and square sum for variance is similar to deriving maximum value, except that the \texttt{FP16} accumulator replaces the comparator.
This process of real-time quantization is pipelined with the dataflow of systolic array.
Therefore, when the last computational dataflow of the systolic array finishes, it only needs several cycles to dequantize and compute the quantization parameter through the RQU.
The RQU improves the efficiency of real-time quantization for activations and the KV cache.

\subsection{Buffers}

We use multi-bank structures for input, weight, output, and quantization buffers to enable parallel access.
\Fig{fig:arch_cmp} shows that the 32 outputs from different rows and columns are dequantized in one cycle. 
It requires 32 scaling factors for activations ($s_X$), 32 for weights ($s_W$), and 32 $a$.
The $s_W$ and $a$ are fixed, so the key point is to ensure that $s_X$ are stored in different banks of quantization buffer to prevent bank conflict.

\subsection{GEMM Computation}
\label{sec:quant_dequant}

We explain how our \proj{}-based accelerator design fuses the (de)quantization process with the original GEMM computation to hide their overhead.


\begin{figure}[t] 
    \centering 
    \includegraphics[width=1\linewidth]{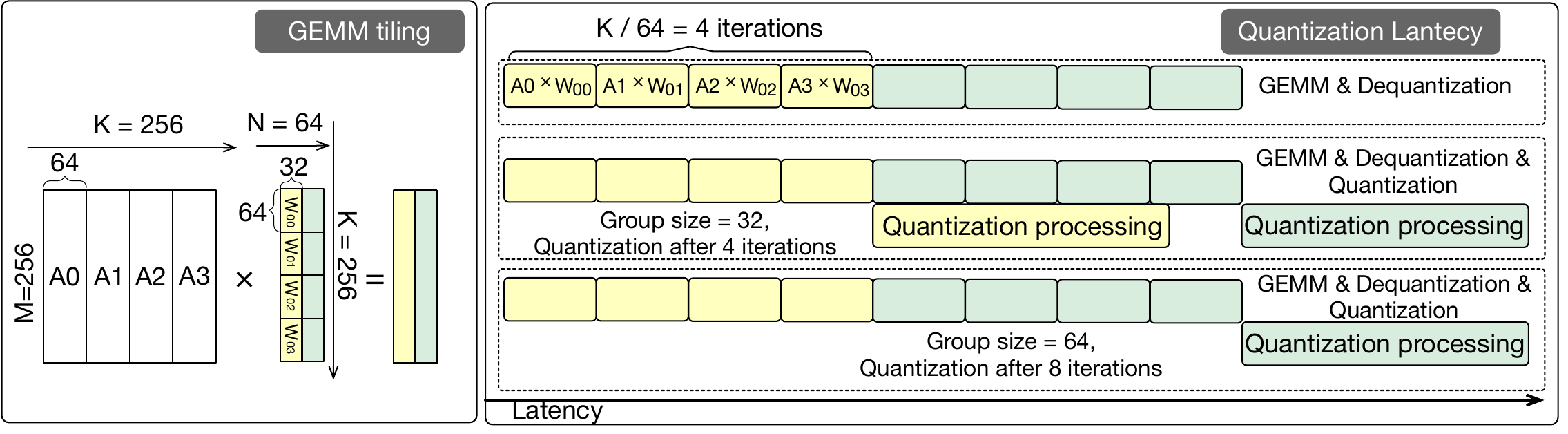}  
    \caption{An example of GEMM tiling and the corresponding quantization overhead.}
    \label{fig:arch_pipeline}
\end{figure}

\paragraph{Dequantization.}
The dequantization process of a group-quantized tensor is to multiply each element with its scaling factor.
Typically, as shown in \Fig{fig:arch_pipeline}, the $M\times K \times N$ GEMM computation is tiled to run on the PE array. 
As long as the group quantization size is greater than the PE array's accumulation dimension (i.e., the $K$ dimension), the scaling factor multiplication can be deferred to after the output of PE array and before the accumulation of the partial sum since all elements within the array share the same scaling factor.
As such, we can concurrently compute the scaling factor product using the vector unit when the PE array starts computation, and augment the original accumulator design with additional multipliers.
Since this accumulation is fully pipelined, its impact on the overall latency is negligible.

\paragraph{Quantization.}
The quantization only occurs when all partial sums finish the accumulation to the global sum. 
In \Fig{fig:arch_pipeline}, after four iterations along the $K$ dimension, the global sum and maximum value for the output tile are determined.
The vector units then compute the scaling factor based on the maximum value and perform division for that output tile.
Note that the first division for the scaling factor occurs once for an entire group, and the second division occurs for each element.
\Fig{fig:arch_pipeline} shows that quantization latency can be hidden by overlapping with GEMM tile computation.
In our design, we model a 12-cycle non-pipelined division unit, which requires  12 $K$-dimension iterations to completely hide this latency.
We show that this has little impact on our overall performance.
In summary, \proj implements the operation fusion~\cite{niu2021dnnfusion,zhou2023ugrapher,zheng2023chimera,zhap2022tacker} between decoding, GEMM, dequantization, and quantization.



\section{Evaluation}
\label{sec:evaluation}

\begin{table*}[t]
    \centering
    \small
    \renewcommand{\arraystretch}{1.2}
    \setlength\extrarowheight{-0.7pt}
    \caption[]{The PTQ results of Wikitext on various LLMs, the metric is perplexity, and the lower is better. ``Act.'' is the activation, ``W'' is the weight, and ``Atten.'' is the attention layer. \proj employs group-wise quantization. ANT, OliVe, and Tender use tensor/channel-wise quantization.
    }
    \resizebox{1.9\columnwidth}{!}{
    \scriptsize
      \begin{tabular}{l|cccc|cccccc|ccccc}
        \Xhline{1.2pt}
        \textbf{Method} & \multicolumn{2}{c}{Linear (bit)} & \multicolumn{2}{c|}{Atten. (bit)} & \multicolumn{6}{c|}{\textbf{LLaMA 1\&2} (Perplexity)}     & \textbf{OPT} & \textbf{OPT} \\  
  
           &  Act. & W & Act. & KV &\textbf{1-7B}  & \textbf{1-13B} & \textbf{1-30B}  & \textbf{1-65B} & \textbf{2-7B} & \textbf{2-13B} & \textbf{6.7B} & \textbf{13B}   \\
  
        \Xhline{1.2pt}
        FP16   & 16 & 16 & 16 & 16 &5.68 & 5.09 & 4.10 & 3.53 & 5.47 & 4.88 & 10.86 & 10.13  \\
        \hline
        \hspace{0.5em} ANT & 4 & 4  &16  & 16 & 61.35 & 53.74 & 19.31 & 21.57  & 124.18 & 333.29 & 6.4E+3  & 9.6E+3 \\
        \hspace{0.5em} OliVe & 4 & 4  &16  & 16 & 32.15 & 15.64 & 13.59  & 12.85  & 44.07 & 50.28 & 39.18  & 65.42 \\
        \hspace{0.5em} Tender & 4 & 4  &16  & 16 & 23.85 & 13.68 & 12.07 & 8.85 & 36.47 & 55.08 & 13.56 & 16.43 \\
        \hspace{0.5em} \proj & 4 & 4  &16  & 16 & \textbf{6.09} & \textbf{5.38} & \textbf{4.42}  & \textbf{3.83} & \textbf{5.92} & \textbf{5.24} & \textbf{11.29} & \textbf{10.62}   \\
        \hline

        \hspace{0.5em} ANT & 8 & 8  &16  & 16 & 9.50  & 8.46  & 6.70  & 5.34  & 10.68 & 16.11 & 14.60  & 165.9 \\
        \hspace{0.5em} OliVe & 8 & 8  &16  & 16  & 5.86  & 5.28  & 4.37  & 3.80  & 5.73 & 5.06 & 11.24  & 10.49  \\
        \hspace{0.5em} Tender & 8 & 8  &16  & 16 & 5.87 & 5.28 & 4.27 & 3.74 & 5.77 & 5.09 & \textbf{10.93} & \textbf{10.17}  \\
        \hspace{0.5em} \proj & 8 & 4  & 16  & 16 & \textbf{5.79}  & \textbf{5.20}  & \textbf{4.18}  & \textbf{3.61}  & \textbf{5.57} & \textbf{4.96} & 10.98  & 10.18  \\
        \hspace{0.5em} \proj & 8 & 4  & 8  & 4 & 5.97 & 5.31 & 4.37 & 3.75 & 5.79 & 5.14 & 11.14 & 10.31  \\ 

        \Xhline{1.2pt}
    \end{tabular}
    }
    \vspace*{0.1cm}

    \label{tbl:ptq_result}
    \vspace*{-0.4cm}
  \end{table*}

We implement the \proj quantization framework and compare its accuracy with four baselines:  Tender~\cite{lee2024tenderacceleratinglargelanguage}, OliVe~\cite{guo2023olive}, ANT~\cite{guo2022ant}, and BitFusion~\cite{sharma2018bit}.
Subsequently, we conduct a detailed analysis of these accelerators' area, performance, and energy consumption, focusing on both the linear and attention layers.
Based on this analysis, we evaluate the performance across various sequence lengths. 
Finally, we evaluate the results by employing group-wise quantization in our baselines.

\subsection{Experimental Setup}
\label{sec:eval_setup}
\paragraph{Models and Datasets. } 
We evaluate \proj on three families of Large Language Models (LLMs), including LLaMA-1 7-65B~\cite{touvron2023llama}, LLaMA-2 7B~\cite{touvron2023llama2}, LLaMA-2 13B, OPT-6.7B~\cite{zhang2022opt} and OPT-13B.
Our evaluation covers different model architectures and sizes.
We evaluate the quantization results using the zero-shot task on the Wikitext dataset (wikitext-2 version)~\cite{merity2016pointer}, with perplexity (PPL) of the generated sentences serving as the metric.
The lower PPL indicates the better results.
We take two subsets from the Pile dataset\mbox{~\cite{gao2020pile}} as the calibration dataset to avoid a biased result.
The OPT and LLaMA models are evaluated by lm-eval-harness framework~\cite{eval-harness}.

We employ \proj on generation tasks to evaluate the KV cache quantization in decode stage.
We use TruthfulQA~\cite{lin2022truthfulqameasuringmodelsmimic} on lm-eval-harness for normal context evaluation and TriviaQA~\cite{joshi2017triviaqalargescaledistantly} on LongBench~\cite{bai2024longbenchbilingualmultitaskbenchmark} for long context evaluation.

\paragraph{Quantization Details. }
We evaluated our approach against three baselines of Transformer quantization accelerators Tender~\cite{lee2024tenderacceleratinglargelanguage}, OliVe~\cite{guo2023olive}, ANT~\cite{guo2022ant} in 4-bit and 8-bit PTQ (Post-Training Quantization) settings.
All of these are open-source frameworks.
ANT performs activation-weight quantization across CNN and Transformer models, such as BERT~\cite{devlin2018bert}, and we extended its framework to support LLM quantization.
Tender and OliVe is the state-of-the-art (SOTA) LLM accelerator that mitigates outliers with a hardware-friendly methodology.

ANT and OliVe apply tensor-wise quantization to activations and channel-wise to weights.
Tender divides activation channels into several chunks. 
In each chunk, Tender further reorders and packages the channels into groups so that the scaling factors of the adjacent groups can be derived via a simple 1-bit shift. 
By moving the shift to accumulation, Tender accommodates outliers with a unified scaling factor in a chunk.
This method is orthogonal to \proj{} and can work together. 
We evaluate Tender here as a SOTA LLM accelerator.
Since Tender, OliVe, and ANT do not quantize the attention layer of LLMs, our analysis of their quantization accuracy is limited to linear layers.

\proj is configured with a group size of 64, applying this setting to quantize activations, weights, and KV cache.
\proj employs \texttt{INT} for activation quantization.
For weight and KV cache quantization, \proj selects the most appropriate encoding from the set $a = \{0, 5, 10, 17, 20, 30, 40, 50, 60, 70, 80, 90, 100, 110, 120\}$ and the \texttt{INT}.
\proj selects the coefficient $a$ for KV cache based on the data variance, as described in \Sec{sec:dse_kv}.

\paragraph{Implementation. }
We evaluate the performance and energy consumption of \proj compared to baseline accelerators.
The \proj's processing elements (PE) and comparator are implemented in Verilog.
They are synthesized using Synopsys' Design Compiler (ver. O-2018.06-SP1) with the TSMC 28nm standard cell library to obtain power and area statistics.
Additionally, the SRAM module's power and area statistics are simulated using CACTI~\cite{muralimanohar2009cacti}.

We develop a simulator based on the cycle-level simulator DNNWeaver~\cite{sharma2016high} to evaluate performance.
ANT, OliVe, and BitFusion, which are built on this simulator, all support mixed precision.
We also implement the architecture of Tender in this simulator and support both \texttt{INT4} and \texttt{INT8} MAC.
ANT has also extended its design using a GPU simulator~\cite{gpgpu-sim2009,leng2013gpuwattch,gpgpu-sim4.0}, but our evaluation focuses exclusively on ASIC implementations.

\paragraph{Accelerator Details. }
BitFusion is the baseline mixed-precision architecture.
ANT is a mixed-precision architecture with adaptive data types.
OliVe further enhances this by supporting outlier-victim pairs with adaptive data types and mixed precision.
Tender can be easily extended to support mixed precision.
Consequently, we consider these architectures as baselines in our evaluation.

For a fair comparison, in the evaluation of the linear layer, we employ the mixed-precision in Tender and OliVe to align the PPL results.
OliVe and Tender utilized 4-8 mixed precision.
The 8-bit ANT can not align the PPL results either, so we label it ANT*.
The 8-bit ANT does not adaptively select the data type and only uses \texttt{INT}, so we can view ANT* as a coarse-grained \texttt{INT8} quantization.

In addition, we configure the number of PEs for each accelerator based on their area overhead.
We aim to compare their performance and power consumption while ensuring nearly equivalent area and PPL.
We set the same configuration for memory bandwidth, on-chip buffer size, and frequency across all accelerators.
In linear layer, the sequence length is set to 2048, with a batch size of 1.

Tender, OliVe, and ANT do not quantize the attention layer in their evaluations.
Therefore, the statistics for these baselines in the attention layer are the same, as shown in \Fig{fig:full_eval}.

\subsection{Accuracy Evaluation}
\label{sec:accuracy}

\paragraph{Results in the Linear Layer. }
In \Tbl{tbl:ptq_result}, we present the PTQ results for the discussed methods.
Both ANT and OliVe implement mixed-precision quantization, but their hardware design requires that activations and weights share the same bit width.
Thus, we evaluate the quantization under 4-bit weight-activation (W4A4) and 8-bit weight-activation (W8A8) settings.
In W4A4 scenario, \proj shows better PPL results than other methods.
\proj with W4A8 achieves under 0.11 PPL loss in LLaMA family models and under 0.12 PPL loss in OPT models.

ANT and OliVe with W4A4 demonstrate significant PPL loss due to the sensitivity of activations, but OliVe's W8A8 configuration shows promising results.
Tender's W4A4 is better than ANT and OliVe because its progressive strategies in the chunk offer a wider region of representation.
Tender's W8A8 result is comparable to OliVe's W8A8.
In OPT models, Tender's W8A8 surpasses \proj's W4A8, while \proj's W4A8 achieves the best perplexity results in the LLaMA family models.


\paragraph{Quantize both the Linear and Attention Layer. }
The Wikitext task only involves the prefill stage, so \Tbl{tbl:ptq_result} reflects the results of quantizing the KV cache during this stage.
Results show that \proj achieves a PPL loss below 0.3 when quantizing both linear and attention layers.
Compared to quantizing only the linear layer, quantizing the attention layer with \proj results in a PPL loss below 0.2.
Quantizing the KV cache along the opposite dimension~\cite{liu2024kivi,hooper2024kvquant} further reduces the loss to 0.1, but performing the low-bit computation in this way is challenging.

\paragraph{Analysis for Generation Tasks.}
For a comprehensive analysis of KV cache quantization, we evaluate \proj on generation tasks TruthfulQA and TriviaQA.
In \Tbl{eval:kv_cache_results}, W4A8 indicates that weights are quantized to 4-bit using \proj, and activations are quantized to \texttt{INT8}.
For KV cache quantization, \proj outperforms \texttt{INT} and achieves a loss below 1.7\%.

\begin{table}[t]
    \centering
    \small
    \renewcommand{\arraystretch}{1.2}
    \caption[]{Evaluation on generation tasks TruthfulQA and TriviaQA. The metric used is the BLEU score for TruthfulQA and the F1 score for TriviaQA. Higher is better.}
    \resizebox{0.8\columnwidth}{!}{
    \begin{tabular}{l|c|ccc}
        \Xhline{1.2pt}
        \textbf{Tasks} & \multicolumn{4}{c}{\textbf{LLaMA-2-7B}}  \\
        \hline
         Weight \& Act. & FP16   & \multicolumn{3}{c}{W4A8}  \\
         \hline
          KV cache & FP16   & FP16 & \texttt{INT4} &  4-bit \proj \\
        \Xhline{1.2pt}
        \hspace{0.5em} TruthfulQA & 27.88 & 27.55 & 25.48 &  26.19  \\
        \hspace{0.5em} TriviaQA & 87.72 & 86.38 & 85.13 & 86.86  \\
        \Xhline{1.2pt}
    \end{tabular}
    }
    \vspace*{0.1cm}
  
    \label{eval:kv_cache_results}
    \vspace*{-0.4cm}
  \end{table}

\begin{table}[b]
    \ra{1.5}
    \caption{The area of core components and buffers for \proj{} and other baseline models using a 28nm process.}
  
    \resizebox{\columnwidth}{!}{%
    \begin{tabular}{c|lcc|c}
    \Xhline{1.2pt}
    \multirow{2}{*}{Arch.} & \multicolumn{3}{c|}{Core}  & \multirow{2}{*}{Others}  \\ \cline{2-4} 
  
    & Component & Number & Area ($mm^2$) &  \\ \Xhline{1.2pt}
    
    \multirow{2}{*}{\proj{}}& 8-bit PE (281.75$\mu m^2$) & 1024 &\multirow{2}{*}{0.302} & Buffer: \\ 
    & RQU (416.63$\mu m^2$) & 32 &  & 512KB, 4.2 $mm^2$ \\ \cline{1-4}
  
    \multirow{3}{*}{OliVe{}} & 4-bit PE (79.57$\mu m^2$) & 4096 &\multirow{3}{*}{0.337}  &   \\ 
    & 4-bit Decoder (48.51$\mu m^2$) & 128 & & Vector Units:  \\ 
    & 8-bit Decoder (73.25$\mu m^2$) & 64 & & \#64, 0.069 $mm^2$\\  \cline{1-4}
  
    \multirow{2}{*}{ANT{}}& 4-bit PE (79.57$\mu m^2$) & 4096 &\multirow{2}{*}{0.327} &   \\ 
    & Decoder (4.9$\mu m^2$) & 128 & & Accumulation Units:\\ \cline{1-4}
  
    Tender & 4-bit PE (77.28 $\mu m^2$) & 4096 & 0.317  & \#32, 0.016 $mm^2$  \\ \cline{1-4}

    \Xhline{1.2pt}
    \end{tabular}%
    }
    \vspace*{1mm}
    \label{tab:area}
  \end{table}

\subsection{Performance, Energy, and Area Evaluation}

\paragraph{Area.}
\Tbl{tab:area} shows the breakdown of the components for \proj and other accelerators, which are all synthesized using a 28~nm process~\cite{guo2022ant,lee2024tenderacceleratinglargelanguage}.
Note that Olive was originally synthesized with 22~nm process~\cite{guo2023olive}.
All accelerators are configured with identical buffer sizes.
In addition, their vector units are the same in our evaluation, so the area is not listed in \Tbl{tab:area}.
In the subsequent evaluation, the number of cores used in the simulator matches those listed in \Tbl{tab:area}.

\begin{figure}[t] 
    \centering 
    \includegraphics[width=0.9\linewidth]{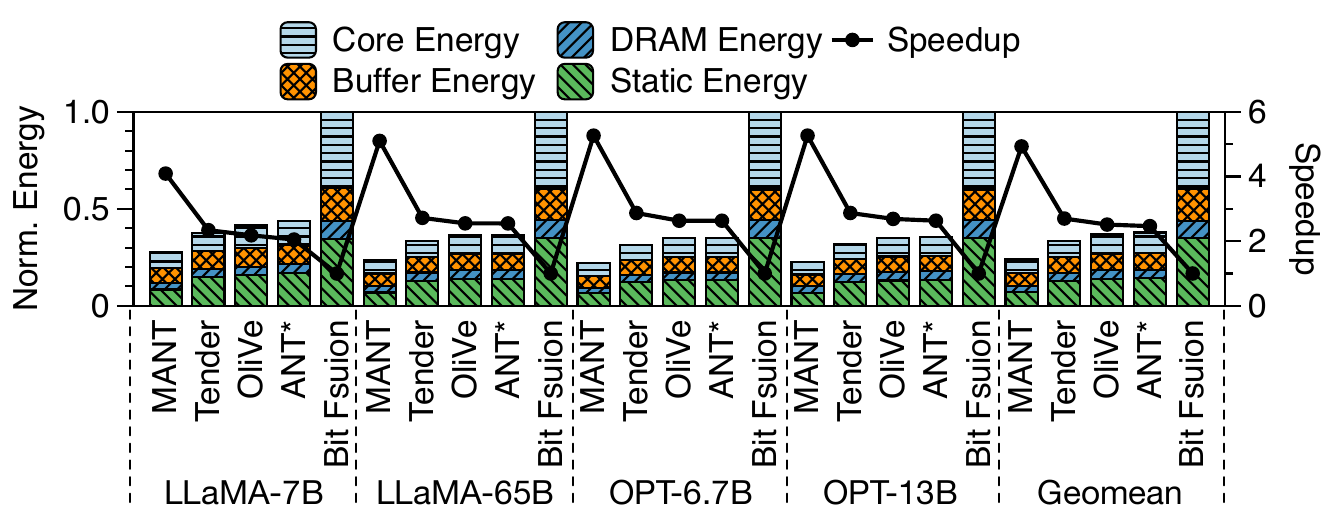}  
    \caption{Speedup and energy breakdown in the linear layer on accelerators. The performance and power consumption of accelerators are evaluated under the nearly equivalent area and PPL except ANT, so it is denoted as ANT*.} 
    \label{fig:lieanr_perf}
  \end{figure}

\paragraph{Linear Layer.}
\Fig{fig:lieanr_perf} presents the performance and energy results of various methods across LLaMA-7B\&65B and OPT-6.7B\&13B.
Given the similarity in LLaMA family models, we focus our evaluation on LLaMA-7B and the larger LLaMA-65B.
Note that ANT* can not recover the accuracy to align with Tender, OliVe, and \proj.
A comparison between OliVe and BitFusion explains the advantages of efficient outlier handling.
BitFusion has higher latency and energy because it performs computation in 8 and 16 bits.
Tender outperforms Olive because the 8-bit layer is less than OliVe.
By leveraging flexible encoding and group-wise quantization, \proj outperforms Tender, OliVe, ANT*, and BitFusion with average speedups of 1.83$\times$, 1.96$\times$, 2.00$\times$, and 4.93$\times$, respectively.

Compared to Tender, OliVe, and ANT*, the energy efficiency of \proj mainly comes from static energy, which is related to execution cycles.
The reduced bit width in \proj lowers DRAM and buffer energy.
The more dequantization in group quantization and the extra shift operation increase the power usage of cores.
Thus, MANT has similar core energy to other baselines, even with a lower bit width.
In total, \proj achieves energy reductions of 1.39, 1.54, 1.57, and 4.16 times compared to Tender, OliVe, ANT*, and BitFusion, respectively.

A typical of GEMM ($M \times K \times N$) in LLaMA-7B is (M, 4096) $\times$ (4096, 4096). 
In this case, the non-overlapped quantization overhead (in the last iterations) occupies 0.3\%. This ratio is decided by the $K$ and $N$, as we detail in \Sec{sec:quant_dequant}.

\begin{figure}[t]
    \centering
    \subfloat[Speedup.]{\label{fig:full_speedup}
    \includegraphics[width=0.42\textwidth]{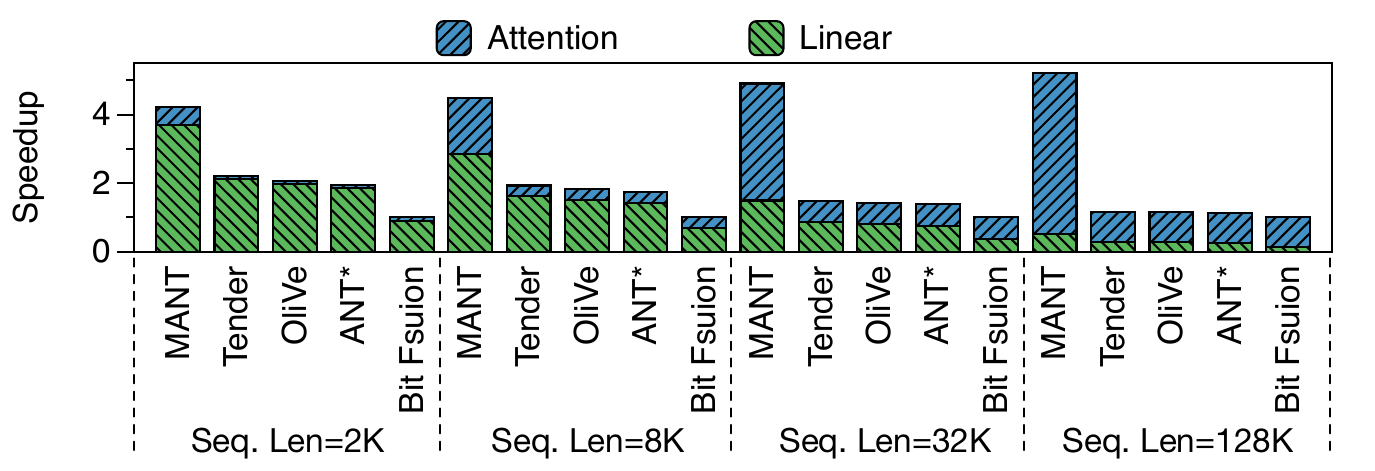}}
  
    \subfloat[Energy.]{
    \label{fig:full_energy}
    \includegraphics[width=0.42\textwidth]{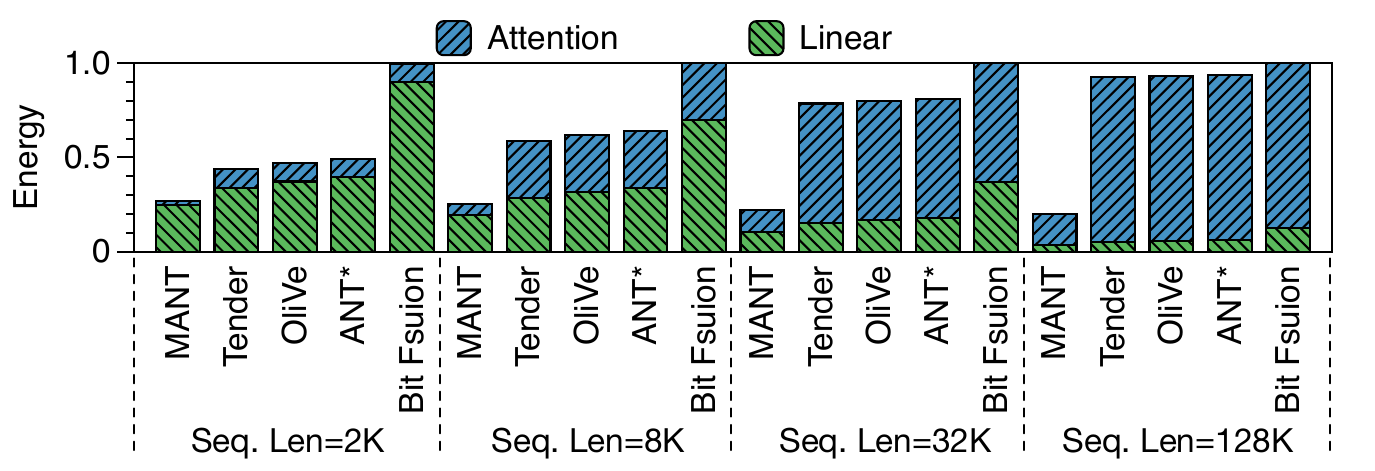}%
    }
    \caption{(a) The speedup of all layers on various accelerators. (b) The normalized energy of all layers on various accelerators.}\label{fig:full_eval}
  \end{figure}

\paragraph{Performance and Energy for All Layers. }
We evaluate the performance and energy of complete layers containing linear and attention.
\Fig{fig:full_eval} shows the speedups \proj achieves over other accelerators at various sequence lengths.
The evaluation uses the LLaMA-7B model with sequence lengths ranging from 2K to 128K.
The baselines do not quantize the attention layer and, therefore, employ 16-bit computation in this layer.

The overall model performance is decided by both linear and attention layers.
As the sequence length increases, the attention layer becomes increasingly dominant.
With 2K sequence length, the linear layer's latency surpasses that of the attention layer, so the overall speedup and energy reduction are primarily determined by the linear layer.
However, when the sequence length reaches 128K, the impact of speedup and energy reduction in the linear layer nearly diminishes.
With a sequence length of 128K, OliVe achieves only a 1.15$\times$ speedup, and Tender achieves a 1.17$\times$ speedup over BitFusion.
Therefore, at a sequence length of 128K, the improvement of \proj is primarily determined by the attention layer.

\proj consistently delivers speedups between 2.04-4.54$\times$ and energy reductions from 1.76-4.12$\times$ compared to OliVe across different sequence lengths.
Moreover, \proj achieves on average 2.99$\times$ (up to 4.46$\times$) speedup and 2.81$\times$ (up to 4.10$\times$) energy reduction to Tender in different sequence lengths.

\begin{figure}[t] 
    \centering 
    \includegraphics[width=0.9\linewidth]{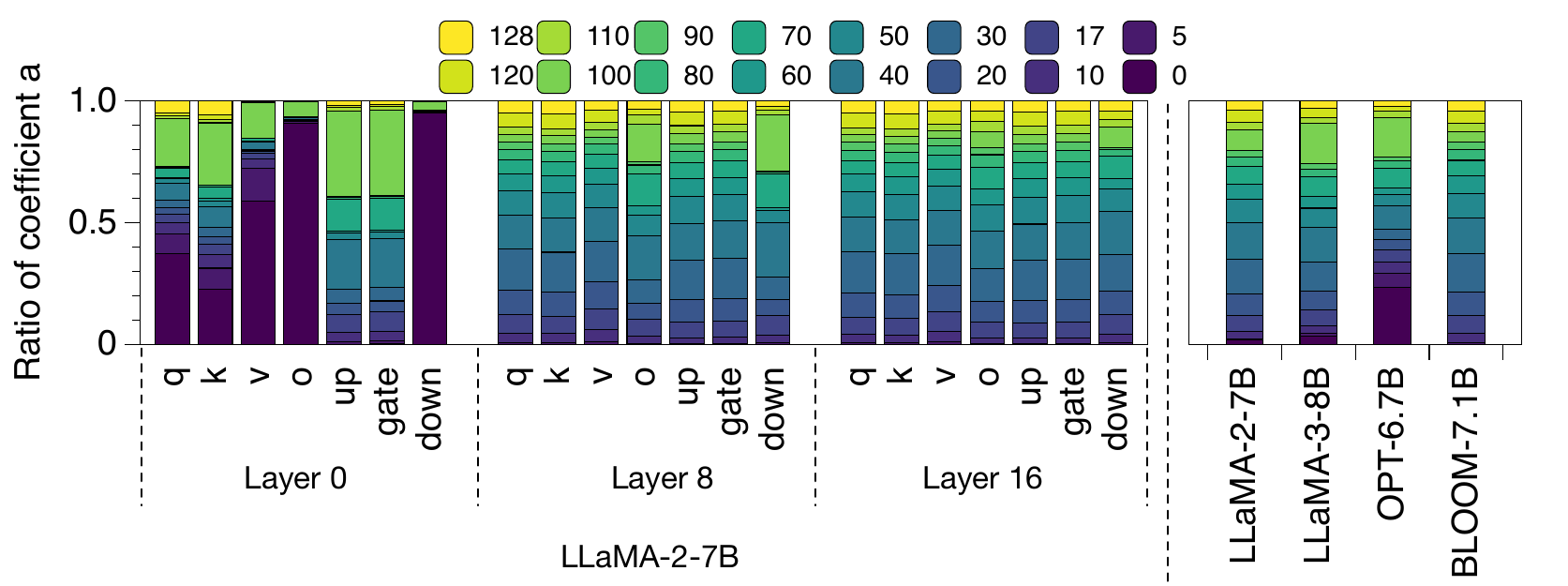}  
    \caption{The data type ratio in various tensor, layer, and models.}
    \label{fig:data_ratio}
  \end{figure}

\paragraph{Data Type Ratio.}
\Fig{fig:data_ratio} shows the selection ratio of $a$ in different tensors, layers, and models.
The $\text{layer}_0$ of LLaMA-2-7B and OPT-6.7B mostly select $a=0$, while other layers and models have a relatively uniform selection.

\begin{figure}[t] 
    \centering 
    \includegraphics[width=0.9\linewidth]{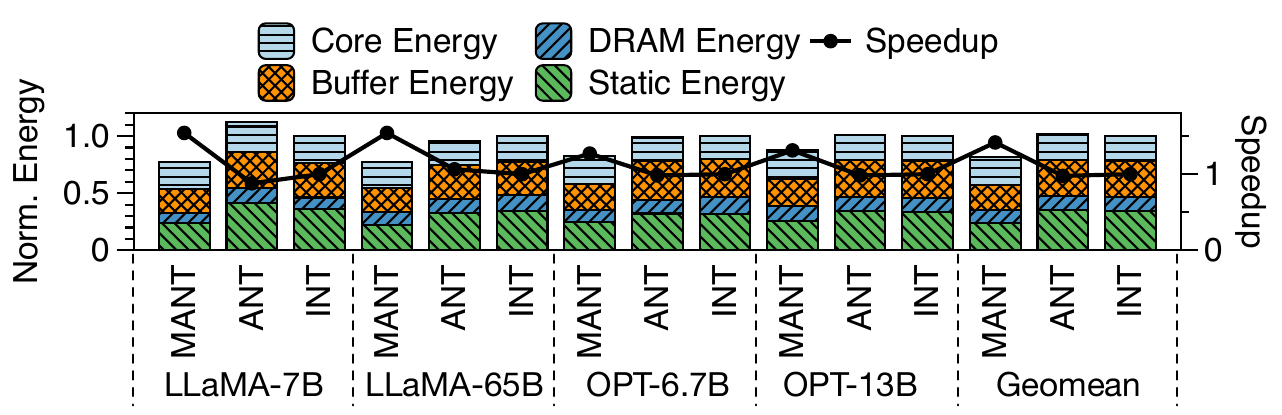}  
    \caption{Speedup and energy breakdown in the linear layer. All quantizations use the group size of 64.} 
    \label{fig:lieanr_group}
  \end{figure}

\subsection{Comparison and Discussion of Group-wise Methods} 
\label{sec:evla_group}

Group-wise quantization can significantly improve the results.
To discuss the benefits of adaptive data types, we compare the perplexity results with group-wise \texttt{INT} quantization.
In addition, we extended ANT and OliVe to support group-wise quantization.
For a fair accuracy comparison, we dynamically calculate the per-group scaling factor in activation quantization for all methods.
It is worth noting that the other methods do not optimize the process of scaling factor computation, leading to non-negligible quantization overhead.
\Tbl{tbl:w4a4_result} shows the group-wise comparison of 4-bit weight and activation quantization.

ANT faces challenges in group-wise quantization.
For weight quantization, ANT can easily select data types from \texttt{INT}, \texttt{flint}, and \texttt{PoT} for each group.
However, ANT does not support real-time data type selection for activation quantization.
It is difficult to determine data types for each group offline due to variable sequence lengths.
Therefore, ANT adaptively selects per-group data types for weight quantization and per-tensor data types for activation, using a per-group scaling factor for both weight and activation.
Selecting a single data type for the entire tensor is typically unsuitable for each group, making ANT less effective than naive \texttt{INT} when the group size is 64 or 32.
OliVe performs better than ANT and \texttt{INT} when the group size is 128.
However, OliVe does not benefit from smaller group sizes.
Olive sacrifices a victim value to provide more bit width for the outlier value.
However, while group-wise quantization mitigates the effect of outliers, as the group size decreases, the negative impact of victims in OliVe outweighs the benefits of protecting outliers.

\proj outperforms group-wise OliVe, ANT, and \texttt{INT} across different group sizes.
Besides, we compare the perplexity with \texttt{MXFP} using a group size of 32.
The scaling factor with an 8-bit exponent (E8M0) increases the quantization error, raising the PPL to 7.16.
The advantage of \texttt{MXFP} is the fast data type casting from \texttt{FP16} to \texttt{MXFP}.

To isolate the benefits of the MANT data type, we extend ANT architecture to support group quantization.
For weights, we determine the data types for each group individually.
Since ANT does not support real-time data type selection, we use a calibration dataset to select a global data type for the entire tensor and apply per-group scaling factors for activations.
Additionally, we implement group-wise \texttt{INT} for real-time KV cache quantization.
To ensure comparable PPL performance with MANT, we use a 4/8 mixed precision configuration for ANT.
As illustrated in \Fig{fig:lieanr_group}, with the same group size of 64, MANT achieves an average speedup of 1.70$\times$ speedup and 1.55$\times$ energy efficiency over ANT.

\begin{table}[t]
    \centering
    \small
    \renewcommand{\arraystretch}{1.2}
    \caption[]{W4A4 results in different group sizes. The dataset is Wikitext. The PPL of \texttt{FP16} is 5.47.}
  
    \begin{tabular}{ll|ccccc}
        \Xhline{1.2pt}
        \multicolumn{2}{c|}{LLaMA-2-7B} & \proj & OliVe & ANT & INT & MXFP4    \\  
  
        \Xhline{1.2pt}
        \multirow{3}{*}{W4A4} & G-128  & 6.26  & 6.43 & 6.49 & 6.54 & - \\
        & G-64   & 5.91 & 6.31 & 6.38 & 6.14 & - \\
        & G-32   & 5.76 & 6.72 & 6.23 & 5.95 & 7.16 \\

        \Xhline{1.2pt}
    \end{tabular}
    \vspace*{0.1cm}
    \label{tbl:w4a4_result}
    \vspace*{-0.4cm}
  \end{table}

\section{Conclusion}
\label{sec:conclusion}
Our paper introduces a new encoding paradigm called \proj, designed to support various data distributions and heterogeneous computability for group-wise weight quantization.
To solve the challenge of employing group-wise quantization in KV cache, we propose real-time quantization components.
We extend the processing unit (PE) in a systolic array to support the efficient computation of \proj, and we implement the real-time update components through the lightweight comparator and fixed-point multiplier.
Our evaluation shows \proj achieves the minimal accuracy loss in several LLMs and achieves on average 2.99$\times$ (up to 4.46$\times$) speedup and 2.81$\times$ (up to 4.10$\times$) energy reduction to the SOTA LLM accelerator.

\section*{Acknowledgment}
This work was supported by the National Key R\&D Program of China under Grant 2022YFB4501400, the National Natural Science Foundation of China (NSFC) grant (62222210, U21B2017 and 62072297). This work was also supported by Shanghai Qi Zhi Institute Innovation Program SQZ202316.  
Cong Guo was supported by Shanghai Jiao Tong University Outstanding Doctoral Graduate Development Scholarship.
The authors express their gratitude to the anonymous reviewers for their insightful feedback, which greatly contributed to improving this work.

\bibliographystyle{IEEEtranS}
\bibliography{paper}

\newpage

\end{document}